\documentclass[final,3p,times]{elsarticle}

\usepackage{amssymb}
\usepackage{amsthm}
\usepackage{amsmath}
\allowdisplaybreaks     
\usepackage{microtype}  
\usepackage{listings}   
\usepackage{xcolor}     
\usepackage{booktabs}   
\usepackage{graphicx}   
\usepackage{dirtree}    
\usepackage{float}      

\biboptions{compress}   

\newcounter{bla}

\definecolor{backcolour}{RGB}{238, 248, 255}
\lstdefinestyle{cpc_listing}{
    basicstyle=\ttfamily\small,  
    backgroundcolor=\color{backcolour},
    columns=fullflexible,
    keepspaces=true,             
    breaklines=true,             
    breakatwhitespace=true,
    prebreak=\mbox{\space\textbackslash}
}

\journal{Computer Physics Communications}

\begin{document}

\begin{frontmatter}

\title{\textls[120]{HELIOS}: A surface integral equation software for light scattering in homogeneous, periodic, and stratified environments}

\author[a]{Parmenion S.~Mavrikakis\corref{author}}
\author[a]{Olivier J.~F.~Martin}

\cortext[author] {Corresponding author.\\\textit{E-mail address:} parmenion.mavrikakis@epfl.ch}
\address[a]{Nanophotonics and Metrology Laboratory (NAM), École Polytechnique Fédérale de Lausanne (EPFL), CH-1015 Lausanne, Switzerland}

\begin{abstract}
We present \textls[120]{HELIOS} (HomogEneous and Layered medIa Optical Scattering), an open-source surface integral equation (SIE) software designed for modeling light scattering by particles embedded in homogeneous or layered media and periodic backgrounds. The code implements the Poggio-Miller-Chang-Harrington-Wu-Tsai (PMCHWT) formulation that has demonstrated exceptional reliability in solving scattering problems with penetrable objects. Domain boundaries are discretized using triangular meshes, upon which the electric and magnetic surface current densities are expanded using the Rao-Wilton-Glisson (RWG) basis functions. For periodic structures, such as photonic crystals and metasurfaces, \textls[120]{HELIOS} employs Ewald's transformation to efficiently evaluate the infinite series associated with 2D lattices. Regarding stratified media, the code utilizes a matrix-friendly approach for the layered media Green's tensor, computing Sommerfeld integrals and accelerating calculations through a tabulation-interpolation scheme. The source code is implemented in C\texttt{++}, while a Python interface manages the workflow, including simulation setup, solver run, and post-processing. The accuracy and versatility of \textls[120]{HELIOS} are demonstrated through various examples that cover all its functionalities.
\\


\noindent \textbf{PROGRAM SUMMARY}

\begin{small}
\noindent
{\em Program Title:} \textls[120]{HELIOS} \\
{\em CPC Library link to program files:} (to be added by Technical Editor) \\
{\em Developer's repository link:} https://github.com/mavrikak/Helios \\
{\em Licensing provisions:} GNU General Public License Version 3 \\
{\em Programming language:} C\texttt{++} and Python \\
{\em Nature of problem:} The program simulates electromagnetic scattering by nanostructures embedded in homogeneous, stratified or periodic backgrounds. It models arbitrary 3D geometries, ranging from isolated scatterers to infinite periodic structures. The code calculates key optical quantities, such as scattering, absorption, and extinction cross-sections, reflectance and transmittance spectra, and near-field distributions, which are essential for analyzing nanophotonic devices and experiments.\\
{\em Solution method:} \textls[120]{HELIOS} uses the SIE method with the PMCHWT formulation and scatterers are discretized via triangular meshes and RWG functions. For periodic lattices, the code applies Ewald's transformation to evaluate the periodic Green's tensor. For stratified media, it uses a matrix-friendly formulation and a tabulation-interpolation acceleration scheme.\\
{\em Additional comments including restrictions and unusual features:} All examples (Sections~\ref{sec:iso_hom},~\ref{sec:per_hom}, and~\ref{sec:iso_layered}) were executed on a standard desktop equipped with a 13th Gen Intel i7-13700 CPU (24 threads, 2.1 GHz) and 32 GB of RAM. The provided high-quality results involve some heavy simulations; however, coarser meshes and larger discretization steps for near-field points can be used to significantly decrease the required computational time and resources. The use of \texttt{tmux} is suggested to facilitate long-running simulations within persistent sessions, ensuring execution continues independently of the user's terminal connection. Finally, the \textls[120]{HELIOS} graphical user interface (GUI) is not presented in this paper, but detailed documentation can be found in the repository.\\
   \\

\end{small}
   \end{abstract}
\end{frontmatter}

\section{Introduction}
\label{sec:introduction}

Computational electromagnetics has become an indispensable tool for the investigation of many phenomena \cite{Sancer_1990, Newman_1984, Yee_1966, Chew_2001, Kern_2012, Quaranta_2018, Achouri_2020, Parmenion_2023}, in both academia and industry. In the field of nanophotonics, it enables the rigorous design and analysis of complex optical systems ranging from isolated to periodic nanostructures \cite{Kottmann_2000, Ekinci_2008, Devilez_2010, Hohenester_2012, Darvill_2013, Hohenester_2014, Yang_2014, Waxenegger_2015, Wang_2016, Monticone_2017, Fruhnert_2017, Bontempi_2017, Suryadharma_2017, Hohenester_2018, Tzarouchis_2018, Sun_2018, Sun_2019, Sideris_2019, Ray_2020, Beutel_2021, Ray_2021, Zerulla_2022, Kruk_2022, Hohenester_2022, Garza_2023, Zerulla_2024, Athanasiou_2024}. Among the available numerical techniques, the SIE method stands out for its efficiency and accuracy in modeling 3D scatterers \cite{Mavrikakis_JOSA_2025}. Unlike volume discretization methods, such as the discrete dipole approximation (DDA) \cite{Purcell_1973, Draine_1988, Martin_1998}, the finite-difference time-domain (FDTD) \cite{Zivanovic_1991, Taflove_1995, Yee_1997, Kunz_2018}, and the finite element method (FEM) \cite{Jin_Volakis, Jin_FEM, Silvester_Ferrari_1996}, SIE formulations require meshing only the boundaries of scattering bodies. This dimensionality reduction significantly lowers the computational burden. Furthermore, the SIE method inherently satisfies the Sommerfeld radiation condition, hence it does not require the implementation of artificial absorbing boundary conditions and allows precise computation of far-field quantities and extreme near-field interactions \cite{Kern_SIE, Mavrikakis_ACES_2025}.

Despite these advantages, the implementation of a robust and versatile SIE solver poses significant challenges, particularly when dealing with penetrable objects and complex backgrounds. For penetrable scatterers in a homogeneous background, the PMCHWT formulation is widely regarded as the standard choice, since it provides stable and accurate solutions \cite{Mavrikakis_ACES_2025}. However, modern nanophotonic devices often consist of periodic arrays, such as photonic crystals and metasurfaces, or structures fabricated in stratified media. Modeling these environments requires specialized Green's functions that are far more complex than their free-space counterparts. For periodic structures, the slow convergence of the periodic dyadic Green's tensor necessitates the employment of acceleration techniques, such as Ewald's transformation \cite{Ewald_1921}. Similarly, for scatterers embedded in stratified media, Green's tensor must account for the reflection and transmission at every interface. This involves the evaluation of Sommerfeld integrals, which are computationally expensive and prone to numerical singularities, requiring robust computation schemes to become practical.

While the theoretical foundations for these problems are well-established, there remains a lack of open-source software that integrates all these functionalities into a single framework. To bridge this gap, we present \textls[120]{HELIOS}, an open-source SIE solver developed for the rigorous simulation of light scattering by penetrable bodies in homogeneous and stratified backgrounds and by periodic structures.

\textls[120]{HELIOS} implements the PMCHWT formulation using RWG functions on triangular meshes. It features a modular architecture with a C\texttt{++} core for intensive operations and a Python interface for simulation management, execution and post-processing. Additionally, it provides a GUI which facilitates the whole simulation process. Key capabilities include the modeling of infinite periodic 2D lattices (with 3D unit cells) via Ewald's summation and the efficient handling of multilayered backgrounds using a matrix-friendly formulation with accelerated Green's function evaluation.

The remainder of this paper is organized as follows. Section~\ref{sec:setup_struct} outlines the software structure, installation, and workflow. Section~\ref{sec:iso_hom} details the formulation for isolated scatterers in homogeneous backgrounds. Section~\ref{sec:per_hom} extends this to periodic structures, while Section~\ref{sec:iso_layered} addresses the challenges of stratified media. Finally, a rich variety of examples are provided for every functionality of the software.

\section{\textls[120]{HELIOS} setup and structure}
\label{sec:setup_struct}

\textls[120]{HELIOS} is designed for Linux environments. Windows users may utilize the Windows Subsystem for Linux (WSL) to establish a fully compatible runtime. The installation process requires the git version-control system. To begin, the \textls[120]{HELIOS} repository is cloned from GitHub:
\begin{lstlisting}[style=cpc_listing]
$ sudo apt install git git-lfs && git lfs install
$ git clone https://github.com/mavrikak/Helios.git
\end{lstlisting}
The project directory contains the source code, build scripts, and example input data. To ensure cross-platform consistency, the code relies on a reproducible environment managed by Miniconda. Hence, the dependencies in \texttt{environment.yml} are installed and the \texttt{Helios} environment is activated:
\begin{lstlisting}[style=cpc_listing]
$ cd Helios
$ conda env create -f environment.yml
$ conda activate Helios
\end{lstlisting}
This establishes the necessary packages, including numerical libraries for data handling and post-processing.

Before execution, the C\texttt{++} core of \textls[120]{HELIOS} must be compiled along with its auxiliary C\texttt{++} \cite{Blitz_pp_1998, Lapack_1999} and Fortran \cite{Quadpack_1983, Amos_1986, Williams_regridpack, Flatau_2012} files. The build system employs a standard \texttt{make} workflow, creating separate directories for applications and compiled objects:
\begin{lstlisting}[style=cpc_listing]
$ mkdir apps
$ mkdir build
$ make clean && make -j
\end{lstlisting}
Successful compilation yields executables in the \texttt{apps/} directory and compiled files in \texttt{build/}. Each simulation is organized structurally in its own subdirectory under \texttt{sim\_data/<SimName>/} and requires two inputs: (i) a \texttt{config.txt} file that specifies the simulation parameters (characteristic examples are provided in the \texttt{sim\_data/} directory of \textls[120]{HELIOS}), and (ii) a \texttt{.mphtxt} or \texttt{.msh} mesh file generated with COMSOL or Gmsh, respectively.

Documentation may be rebuilt locally to reflect source code modifications or functional extensions. The documentation build system, located in \texttt{docs/}, utilizes Doxygen to generate HTML and PDF references:
\begin{lstlisting}[style=cpc_listing]
$ cd docs
$ bash build.sh
\end{lstlisting}
Overall, completing these steps ensures a fully configured environment for compiling the project's code and running electromagnetic simulations. In Table~\ref{tab:structure} the main classes of the source C\texttt{++} code are presented.
\begin{table}[h]
    \centering
    \caption{List of selected C\texttt{++} classes. They appear categorized according to their main contribution to the source code.}
    \label{tab:structure}
    \resizebox{\textwidth}{!}{%
        \begin{tabular}{lllllll}
            \toprule
            \textbf{Mesh and} & \textbf{Simulation} & \textbf{Green's} & \textbf{SIE} & \textbf{Incident} & \textbf{Quadrature and} & \textbf{Driver and} \\
            \textbf{Geometry} & \textbf{domains} & \textbf{functions} & \textbf{formulations} & \textbf{fields} & \textbf{Singularity handling} & \textbf{Jobs} \\
            \midrule
            Bsp3D & Domain & GreenF & SIEForm & Dipole & GaussQuad $\langle$Type$\rangle$ & CLParser \\
            RWGFun & DomainHom3D & GreenFHom3D & SIEFormPMCHW & Gaussian & SingCancellation & CLOption $\langle$T$\rangle$ \\
            SurfaceMesh & DomainHom3DPer & GreenFHom3DPer & & IncidentField & SingSub & CLStringOption \\
            Triangle & DomainLayered3D & GreenFHom3DPer2D & & PlaneWave & TriQuadPol & CLFlag \\
            & LayeredMediaUtils & GreenFLayered3D & & & & JobParser \\
            & & LookUpTable & & & & SimJob \\
            & & LookUpTableBin & & & & \\
            & & SymmetryOperation & & & & \\
            & & Translation & & & & \\
            & & SommerfeldIntegrator & & & & \\
            \bottomrule
        \end{tabular}%
    }
\end{table}
A very detailed presentation of its structure and files is provided in the Doxygen documentation. Finally, the project's main structure is presented below:
\dirtree{%
.1 Helios/.
.2 docs/ \DTcomment{C++ code documentation and GUI user's guide}.
.2 include/ \DTcomment{header and blitz++ files}.
.2 materials/ \DTcomment{materials epsilon tables}.
.2 pytools/ \DTcomment{project tools}.
.3 drude\_model.py \DTcomment{Drude model}.
.3 jobwriter.py \DTcomment{creates jobfiles}.
.3 meshconvert.py \DTcomment{creates .mesh}.
.3 plot\_mesh.py \DTcomment{plots mesh}.
.3 visualization.py \DTcomment{plots results}.
.2 sim\_data/ \DTcomment{user data inputs}.
.3 $\langle$SimName$\rangle$/ \DTcomment{simulation data folder}.
.4 config.txt \DTcomment{configuration file}.
.4 mesh.mphtxt \DTcomment{COMSOL mesh file}.
.4 mesh.msh \DTcomment{Gmsh mesh file}.
.2 sim\_res/ \DTcomment{simulations outputs}.
.3 $\langle$SimName$\rangle$/ \DTcomment{results folder}.
.4 jobs/ \DTcomment{job files}.
.4 logs/ \DTcomment{log files}.
.4 out/ \DTcomment{data outputs}.
.5 csc/ \DTcomment{cross-sections}.
.5 fields/ \DTcomment{field files}.
.5 media/ \DTcomment{PNG plots}.
.4 points/ \DTcomment{points files}.
.4 lambdalist.txt \DTcomment{wavelength list}.
.4 mesh.mesh \DTcomment{converted mesh file}.
.2 src/ \DTcomment{C++ source code}.
.2 third\_party/ \DTcomment{Fortran files}.
.3 amos/ \DTcomment{amos library files}.
.3 legacy\_f/ \DTcomment{iterative solver}.
.3 quadpack/ \DTcomment{quadpack routines}.
.3 regridpack/ \DTcomment{regridpack routines}.
.3 erfc.bin \DTcomment{erfc lookup table}.
.2 dist\_run\_helios.py \DTcomment{parallel job launcher}.
.2 environment.yml \DTcomment{conda environment}.
.2 helios\_gui.py \DTcomment{GUI launcher}.
.2 makefile \DTcomment{compilation instructions}.
.2 README.md \DTcomment{README file}.
.2 run\_sie.py \DTcomment{code run script}.
}

\section{Isolated scatterers in homogeneous backgrounds}
\label{sec:iso_hom}

In this section, we present the theoretical formulation and numerical implementation for modeling isolated scatterers in homogeneous media. Following the theoretical overview, we provide two step-by-step examples demonstrating how to set up, run, and analyze simulations. For the rest of this work, time-harmonic fields with a $\exp(-j \omega t)$ time dependence are considered.

\subsection{Theory}
\label{subsec:iso_hom_theory}

\begin{figure}[htbp]
\centering\includegraphics[width=0.95\columnwidth]{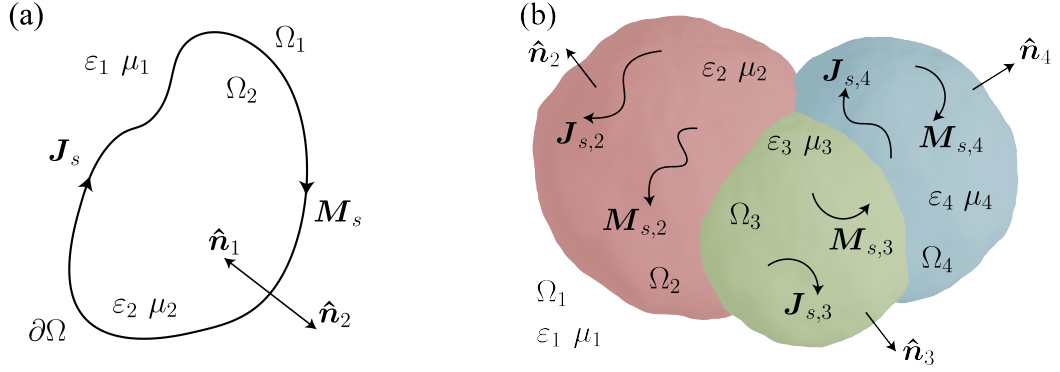}
\caption{Penetrable scatterers in a homogeneous medium. (a) Single particle scenario (domain $\Omega_2$), and (b) multiple attached particles case (domains $\Omega_2$, $\Omega_3$, $\Omega_4$). In both cases the homogeneous background domain is $\Omega_1$.}
\label{fig:domains_hom}
\end{figure}

Let us focus on the two-region scattering problem illustrated in Fig.~\ref{fig:domains_hom}(a), which consists of a homogeneous body embedded in a homogeneous background. Each region $\Omega_i$ is characterized by its electric permittivity and magnetic permeability pair $(\varepsilon_i,\mu_i)$. The fields satisfy the standard vector Helmholtz equations,
\begin{align}
    \nabla \times \nabla \times \mathbf{E}_i(\mathbf{r})-k_i^2 \mathbf{E}_i(\mathbf{r})&=j \omega \mu_i \mathbf{J}_i(\mathbf{r}) \,,
    \label{eq:double_curl_Ei} \\
    \nabla \times \nabla \times \mathbf{H}_i(\mathbf{r})-k_i^2 \mathbf{H}_i(\mathbf{r})&=\nabla \times \mathbf{J}_i(\mathbf{r}) \,,
    \label{eq:double_curl_Hi}
\end{align}
where $\omega$ denotes the angular frequency, $k_i = \omega \sqrt{\varepsilon_i \mu_i}$ symbolizes the wavenumber in $\Omega_i$ and $\mathbf{J}_i(\mathbf{r})$ is the volume current distribution in $\Omega_i$.
By introducing the dyadic Green's tensor
\begin{equation}
    \overline{\mathbf{G}}\left(\mathbf{r}, \mathbf{r}^{\prime}\right)=\left(\overline{\mathbf{1}}+\frac{\nabla \nabla}{k^{2}}\right)\frac{\exp(j k \left|\mathbf{r}-\mathbf{r^{\prime}}\right|)}{4\pi \left|\mathbf{r}-\mathbf{r^{\prime}}\right|} \,,
    \label{eq:dyadic_G}
\end{equation}
where $g \left(\mathbf{r}, \mathbf{r}^{\prime}\right) = \exp(j k \left|\mathbf{r}-\mathbf{r^{\prime}}\right|)/{4\pi \left|\mathbf{r}-\mathbf{r^{\prime}}\right|}$ is the scalar Green's function in a homogeneous medium with wavenumber $k$, and the following equation \cite{Kern_SPIE} in each domain $\Omega_i$
\begin{equation}
    \nabla \times \nabla \times \overline{\mathbf{G}}_i\left(\mathbf{r}, \mathbf{r}^{\prime}\right)-k_i^2 \overline{\mathbf{G}}_i\left(\mathbf{r}, \mathbf{r}^{\prime}\right)=\overline{\mathbf{1}} \delta\left(\mathbf{r}-\mathbf{r}^{\prime}\right) \,,
    \label{eq:double_curl_Gi}
\end{equation}
where we have Kronecker's delta and $\left. \overline{\mathbf{1}} \right|_{\xi_1\xi_2} = \delta_{\xi_1\xi_2}$, one obtains the familiar electric field integral equation (EFIE) and magnetic field integral equation (MFIE) \cite{Kern_SIE}. For each domain, they read respectively \cite{Mavrikakis_JOSA_2025, Mavrikakis_ACES_2025}
\begin{align}
    j \omega \mu_i \left.\mathbf{\mathcal{D}}_i\left\{\mathbf{J}_{s,i} \right\} \left(\mathbf{r}\right) \right|_{\tan} + \left.\mathbf{\mathcal{K}}_i\left\{\mathbf{M}_{s,i} \right\}\left(\mathbf{r}\right) \right|_{\tan} &= \left.\mathbf{E}_i^{\mathrm{inc}}\left(\mathbf{r}\right) \right|_{\tan} \,, \label{eq:EFIE} \\
    j \omega \varepsilon_i \left.\mathbf{\mathcal{D}}_i\left\{\mathbf{M}_{s,i} \right\} \left(\mathbf{r}\right) \right|_{\tan} - \left.\mathbf{\mathcal{K}}_i\left\{\mathbf{J}_{s,i} \right\}\left(\mathbf{r}\right) \right|_{\tan} &= \left.\mathbf{H}_i^{\mathrm{inc}}\left(\mathbf{r}\right) \right|_{\tan} \,, \label{eq:MFIE}
\end{align}
where ``$\tan$'' stands for tangential, $\mathbf{J}_{s,i}\left(\mathbf{r}\right)$ is the electric surface current density and $\mathbf{M}_{s,i}\left(\mathbf{r}\right)$ is the magnetic surface current density defined on the boundary $\partial \Omega_i$ of domain $\Omega_i$. The depicted linear integral operators $\mathbf{\mathcal{D}}_i\left\{\mathbf{\mathcal{X}} \right\}\left(\mathbf{r}\right)$ and  $\mathbf{\mathcal{K}}_i\left\{\mathbf{\mathcal{X}} \right\}\left(\mathbf{r}\right)$ can be expressed as \cite{Mavrikakis_JOSA_2025, Mavrikakis_ACES_2025}
\begin{align}
    \mathbf{\mathcal{D}}_i\left\{\mathbf{\mathcal{X}} \right\}\left(\mathbf{r}\right) &= \int_{\partial \Omega_i} \mathrm{~d} S^{\prime} \overline{\mathbf{G}}_i\left(\mathbf{r}, \mathbf{r}^{\prime}\right) \cdot \mathbf{\mathcal{X}}\left(\mathbf{r}^{\prime}\right) \,, \label{eq:IO_D}\\
    \mathbf{\mathcal{K}}_i\left\{\mathbf{\mathcal{X}} \right\}\left(\mathbf{r}\right) &= \int_{\partial \Omega_i} \mathrm{~d} S^{\prime} \left[\nabla^{\prime} \times \overline{\mathbf{G}}_i\left(\mathbf{r}, \mathbf{r}^{\prime}\right)\right] \cdot \mathbf{\mathcal{X}}\left(\mathbf{r}^{\prime}\right)\,. \label{eq:IO_K}
\end{align}

By discretizing the scatterer's surface with a triangular mesh $\mathcal{M}$ consisting of $N_e$ edges, expanding the unknown surface current densities of Eqs.~\eqref{eq:EFIE} and~\eqref{eq:MFIE} with the RWG basis functions \cite{Mavrikakis_JOSA_2025, Kolundzija_1999, Oijala_2005} as
\begin{align}
    \mathbf{J}_{s,i}\left(\mathbf{r}\right) &= \sum_{\ell = 1}^{N_e} \alpha_{\ell} \mathbf{f}_{\ell,i}\left(\mathbf{r}\right) \,,\label{eq:Js_discrete} \\
    \mathbf{M}_{s,i}\left(\mathbf{r}\right) &= \sum_{\ell = 1}^{N_e} \beta_{\ell} \mathbf{f}_{\ell,i}\left(\mathbf{r}\right) \,, \label{eq:Ms_discrete}
\end{align}
and applying the Galerkin testing procedure \cite[chap.~3]{Harrington_1993}, the EFIE and MFIE become respectively
\begin{align}
    \left[ j \omega \mu_i \mathbf{D}_i \quad \mathbf{K}_i \right] \cdot \left[\begin{array}{c}
    \boldsymbol{\alpha} \\
    \boldsymbol{\beta}
    \end{array}\right] &= \mathbf{q}_i^E \,,
    \label{eq:EFIE_discrete} \\
    \left[ -\mathbf{K}_i \quad j \omega \varepsilon_i \mathbf{D}_i \right] \cdot \left[\begin{array}{c}
    \boldsymbol{\alpha} \\
    \boldsymbol{\beta}
    \end{array}\right] &= \mathbf{q}_i^H \,,
    \label{eq:MFIE_discrete}
\end{align}
where $\mathbf{D}_i$, $\mathbf{K}_i$ are square matrices of side $N_e$, $\boldsymbol{\alpha}$, $\boldsymbol{\beta}$ are the column vectors that contain the expansion coefficients, and $\mathbf{q}_i^E$, $\mathbf{q}_i^H$ are the electric and magnetic incident field column vectors. The elements of the aforementioned quantities are written as
\begin{align}
    (D_i)_{\xi_1\xi_2} = {}&\int_{\partial \Omega_i} \mathrm{~d} S \mathbf{f}_{\xi_1,i}(\mathbf{r}) \cdot \int_{\partial \Omega_i} \mathrm{~d} S^{\prime} \overline{\mathbf{G}}_i\left(\mathbf{r}, \mathbf{r}^{\prime}\right) \cdot \mathbf{f}_{\xi_2,i}\left(\mathbf{r}^{\prime}\right) \label{eq:Dmn} \,, \\
    (K_i)_{\xi_1\xi_2} = {}&\int_{\partial \Omega_i} \mathrm{~d} S \mathbf{f}_{\xi_1,i}(\mathbf{r}) \cdot \int_{\partial \Omega_i} \mathrm{~d} S^{\prime}\left[\nabla^{\prime} \times \overline{\mathbf{G}}_i\left(\mathbf{r}, \mathbf{r}^{\prime}\right)\right] \cdot \mathbf{f}_{\xi_2,i}\left(\mathbf{r}^{\prime}\right) \label{eq:Kmn} \,, \\
    (q_{i}^E)_{\xi_1} = {}&\int_{\partial \Omega_i} \mathrm{~d} S \mathbf{f}_{\xi_1,i}(\mathbf{r}) \cdot \mathbf{E}_i^{\mathrm{inc}}(\mathbf{r}) \label{eq:Rem} \,, \\
    (q_{i}^H)_{\xi_1} = {}&\int_{\partial \Omega_i} \mathrm{~d} S \mathbf{f}_{\xi_1,i}(\mathbf{r}) \cdot \mathbf{H}_i^{\mathrm{inc}}(\mathbf{r}) \label{eq:Rmm} \,, \\
    \boldsymbol{\alpha} = {}&[\alpha_1 \quad ... \quad \alpha_{N_e}]^T \label{eq:alpha_l} \,, \\
    \boldsymbol{\beta} = {}&[\beta_1 \quad ... \quad \beta_{N_e}]^T \label{eq:beta_l} \,.
\end{align}
However, both the EFIE and MFIE suffer from internal resonances, which produce inaccurate results \cite{Harrington_1989, Sheng_1998}. To overcome this difficulty, the PMCHWT scheme \cite{Poggio_1973, Chang_1977, Wu_1977} combines the equations from all regions into a single coupled system \cite{Kern_SIE}
\begin{equation}
    \sum_i\left[\begin{array}{cc}
    j \omega \mu_i \mathbf{D}_i & \mathbf{K}_i \\
    -\mathbf{K}_i & j \omega \varepsilon_i \mathbf{D}_i
    \end{array}\right] \cdot
    \left[\begin{array}{c}
    \boldsymbol{\alpha} \\
    \boldsymbol{\beta}
    \end{array}\right]
    = \sum_i\left[\begin{array}{l}
    \mathbf{q}_i^E \\
    \mathbf{q}_i^H
    \end{array}\right] \,,
    \label{eq:TPMCHWT}
\end{equation}
thus providing a stable and reliable formulation for scattering problems with multiple penetrable scattering bodies in homogeneous backgrounds, as shown in Fig.~\ref{fig:domains_hom}(b).

Finally, the singularities of the Green's function that occur during numerical integration are handled using the singularity subtraction technique, as described in \cite{HanninenSingSub}. This method requires the Taylor series expansion of Green's function within domain $\Omega_i$, which can then be decomposed into smooth and singular components \cite{Kottmann_VIE} as follows
\begin{equation}
    g_i\left(\mathbf{r},\mathbf{r}^{\prime}\right)=g_i^{sm}\left(\mathbf{r},\mathbf{r}^{\prime}\right)+g_i^{sn}\left(\mathbf{r},\mathbf{r}^{\prime}\right) \,,
\end{equation}
where $g_i^{sm}(\mathbf{r},\mathbf{r}^{\prime})$ is suitable for numerical integration, while $g_i^{sn}(\mathbf{r},\mathbf{r}^{\prime})$ is evaluated semi-analytically using closed-form expressions. Detailed formulas for these calculations with RWG functions are provided in \cite{HanninenSingSub}.

\subsection{Simulating nanoparticles in homogeneous backgrounds}
\label{subsec:iso_hom_simulation}

To illustrate the core workflow of \textls[120]{HELIOS} and establish a  performance baseline, we present two scenarios of light scattering in homogeneous backgrounds (\texttt{----mode 0} is the corresponding command line argument). These configurations facilitate direct comparison with analytical solutions or well-established benchmarks, serving as an ideal validation of the solver's accuracy and post-processing capabilities. Throughout this paper, the following input arguments are used: (i) \texttt{-l 2} to solve the final system with the direct lower-upper (LU) factorization, \texttt{-th 0}  to automatically utilize the maximum available threads on the host machine, and \texttt{-a} to increase the number of numerical integration points per mesh triangle \cite{Raziman_2015}.

\subsubsection{Homogeneous gold sphere in vacuum}
\label{subsubsec:Au_sphere_hom}

\begin{figure}[H]
\centering\includegraphics[width=0.95\columnwidth]{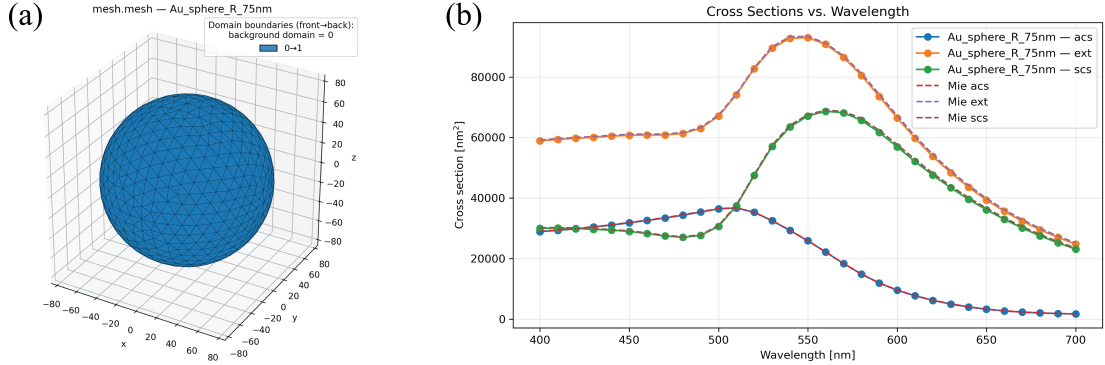}
\caption{Au sphere ($R = 75~\mathrm{nm}$) in free-space. (a) Triangular mesh, and (b) cross-section results comparison between the SIE method and Mie theory for an incident plane wave that propagates towards $+z$ and is $x$-polarized.}
\label{fig:sphere_mesh_cs}
\end{figure}

We first examine the scattering response of a spherical Au nanoparticle in free-space, with radius $R = 75~\mathrm{nm}$. This benchmark system is routinely employed in nanophotonics and serves as a fundamental test of the program's accuracy \cite{Kern_2010}. The simulation is initialized using the \texttt{prepare} command of the \texttt{run\_sie.py} script, which generates the results folder under \texttt{sim\_res/Au\_sphere\_R\_75nm/} with all the required data:
\begin{lstlisting}[style=cpc_listing]
$ python3 run_sie.py prepare Au_sphere_R_75nm --mode 0 --materials eps --spline
\end{lstlisting}
If an earlier version exists, it may be overwritten with:
\begin{lstlisting}[style=cpc_listing]
$ python3 run_sie.py prepare Au_sphere_R_75nm --mode 0 --materials eps --spline --overwrite
\end{lstlisting}
Prior to execution, it is beneficial to inspect the generated mesh. This can be done interactively:
\begin{lstlisting}[style=cpc_listing]
$ python3 pytools/plot_mesh.py --sim Au_sphere_R_75nm --show
\end{lstlisting}
or by saving an image, see Fig.~\ref{fig:sphere_mesh_cs}(a), to the results directory:
\begin{lstlisting}[style=cpc_listing]
$ mkdir sim_res/Au_sphere_R_75nm/out/media/
$ python3 pytools/plot_mesh.py --sim Au_sphere_R_75nm --save ./sim_res/Au_sphere_R_75nm/out/media/mesh.png
\end{lstlisting}
The \texttt{----show} and \texttt{----save} options are available for all visualizations. For the remainder of this work, we will focus on commands that save PNG images. Once the simulation's geometry and settings are validated, the SIE solver assembles and solves the final linear system:
\begin{lstlisting}[style=cpc_listing]
$ python3 run_sie.py solve Au_sphere_R_75nm --mode 0 -l 2 -th 0 -a
\end{lstlisting}
The desired illumination conditions are defined in \texttt{config.txt}.

Key quantities of interest for spherical nanoparticles include the scattering ($C_{scs}$), absorption ($C_{acs}$), and extinction ($C_{ext}$) cross-sections. These can be illustrated in PNG format:
\begin{lstlisting}[style=cpc_listing]
$ python3 pytools/visualization.py Au_sphere_R_75nm --root sim_res --plot-csc --subjob 1 --csc scs --csc acs --csc ext --save ./sim_res/Au_sphere_R_75nm/out/media/cross_sections.png
\end{lstlisting}
To validate against analytical solutions, \textls[120]{HELIOS} allows the overlay of corresponding Mie theory results using identical radii and tabulated material properties, as depicted in Fig.~\ref{fig:sphere_mesh_cs}(b):
\begin{lstlisting}[style=cpc_listing]
$ python3 pytools/visualization.py Au_sphere_R_75nm --root sim_res --plot-csc --subjob 1 --csc scs --csc acs --csc ext --mie-radius 75 --mie-material Au --mie-background eps=1,0 --mie-spline --materials-root materials --save ./sim_res/Au_sphere_R_75nm/out/media/cross_sections_with_Mie.png
\end{lstlisting}

\begin{figure}[H]
\centering\includegraphics[width=0.95\columnwidth]{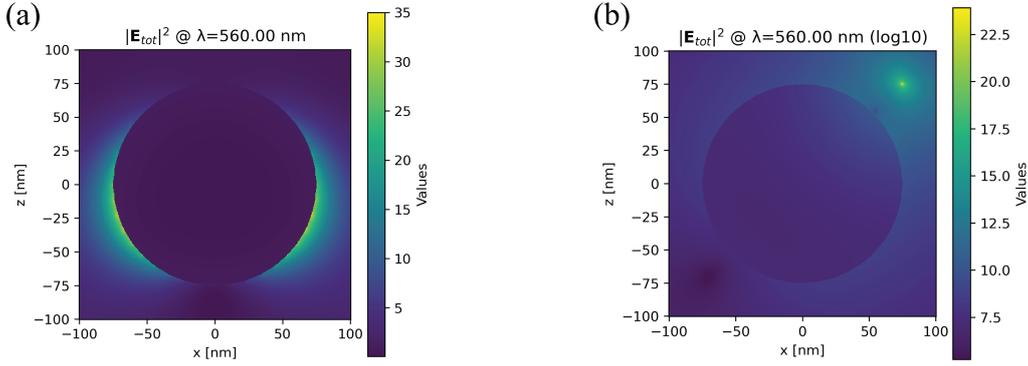}
\caption{Near-field for (a) an incident plane wave that propagates towards $+z$ and is $x$-polarized, and (b) an in-plane source dipole at ($x = 75~\mathrm{nm},\, z = 75~\mathrm{nm}$) with direction ($\theta = \pi/4,\, \varphi = \pi$). The dipole orientation is defined in the standard spherical coordinates.}
\label{fig:sphere_near-fields}
\end{figure}

To visualize near-field, a set of sampling points is required. Here, the points file \texttt{points\_xz.pos} is created for a 2D slice in the $y = 0$ plane, covering the region $x,\, z \in [-100,\, 100]~\mathrm{nm}$ with a step size of 0.5 nm:
\begin{lstlisting}[style=cpc_listing]
$ python3 run_sie.py make-points Au_sphere_R_75nm xz --x -100 100 --z -100 100 --step 0.5 --y0=0 -o points_xz.pos
\end{lstlisting}
Field values at these locations are computed during post-processing for the resonant wavelength of $560~\mathrm{nm}$:
\begin{lstlisting}[style=cpc_listing]
$ python3 run_sie.py post Au_sphere_R_75nm -th 0 -a --lambdas 560 -p points/points_xz.pos
\end{lstlisting}

Finally, two near-field distributions at the sampling points are visualized in Fig.~\ref{fig:sphere_near-fields}. The latter presents the total electric field intensity in the \(xz\)-plane for plane wave ($\mathbf{k}=k_z\hat{\mathbf{z}}$, with $k_z>0$, and $\mathbf{E}^\mathrm{inc} = E^{inc}_x \hat{\mathbf{x}}$) and dipole ($x = 75~\mathrm{nm},\, z = 75~\mathrm{nm},\, \theta = \pi/4,\, \varphi = \pi$) illuminations:
\begin{lstlisting}[style=cpc_listing]
$ python3 pytools/visualization.py Au_sphere_R_75nm --root sim_res --points points_xz.pos --field e --part tot --quantity total --function intensity --mode 2d --subjob 1 --lambda 560 --save ./sim_res/Au_sphere_R_75nm/out/media/PW_560nm_Etot_intensity.png
$ python3 pytools/visualization.py Au_sphere_R_75nm --root sim_res --points points_xz.pos --field e --part tot --quantity total --function intensity --mode 2d --subjob 2 --log --lambda 560 --save ./sim_res/Au_sphere_R_75nm/out/media/dip_560nm_Etot_intensity.png         
\end{lstlisting}
The dipolar source implementation can be useful when studying fluorescence since it can model fluorescent molecules \cite{Girard_2015, Athanasiou_2025}. Treating a fluorophore as a dipole helps explain key phenomena such as polarization effects, emission patterns, energy transfer, and how fluorescence depends on molecular orientation relative to the excitation light.

\subsubsection{$\varphi$-type Janus ring}
\label{subsubsec:phi_type_Janus}

\begin{figure}[htbp]
\centering\includegraphics[width=0.95\columnwidth]{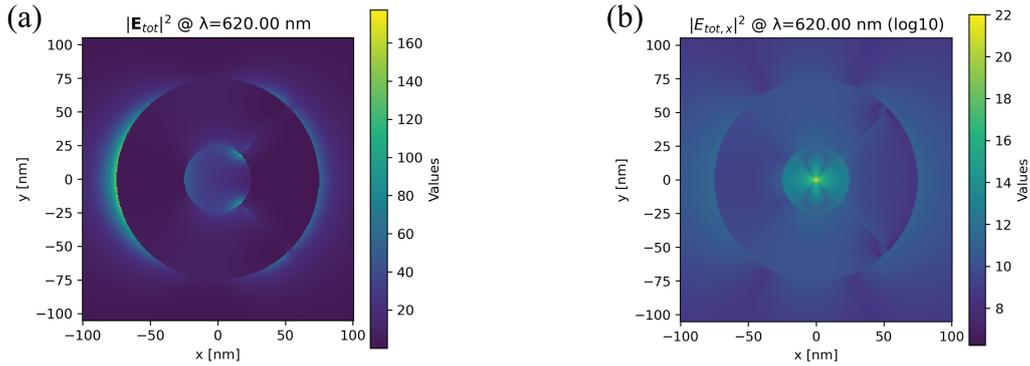}
\caption{Near-field for (a) an incident plane wave that propagates towards $+z$ and is $x$-polarized, and (b) a dipole at the origin with direction ($\theta = \pi/2,\, \varphi = 0$).}
\label{fig:Janus_near-fields}
\end{figure}

Next, we consider an inhomogeneous toroidal nanoparticle wherein the permittivity varies along the azimuthal direction $\varphi$. In contrast to the homogeneous Au sphere, the Janus ring \cite{Honegger_2010, Zentgraf_2010, Garapati_2017} demonstrates the flexibility of \textls[120]{HELIOS} in handling different materials and discretizing junctions within the SIE framework. The simulation setup follows the standard workflow. The results directory is first generated using the \texttt{prepare} command:
\begin{lstlisting}[style=cpc_listing]
$ python3 run_sie.py prepare phi_type_Janus_ring --mode 0 --materials eps --spline
\end{lstlisting}
A visual inspection of the mesh is recommended to verify that the Janus ring geometry and the azimuthally varying material patches are modeled correctly:
\begin{lstlisting}[style=cpc_listing]
$ mkdir sim_res/phi_type_Janus_ring/out/media/
$ python3 pytools/plot_mesh.py --sim phi_type_Janus_ring --save ./sim_res/phi_type_Janus_ring/out/media/mesh.png
\end{lstlisting}
In Fig.~\ref{fig:Janus_mesh_cs}(a), the crystalline Silicon (cSi) section ($\varphi \in [-\pi/4,\, \pi/4]$) is shown in brown, while the Au domain is depicted in blue. Additionally, the interfaces between these domains are highlighted in light blue. The minor radius is $R_{1} = 25~\mathrm{nm}$, while the major radius is $R_{2} = 75~\mathrm{nm}$. Upon geometrical confirmation, the system is assembled and solved for an incident plane wave that propagates towards $+z$ and is $x$-polarized, and a dipole at the origin with direction ($\theta = \pi/2,\, \varphi = 0$):
\begin{lstlisting}[style=cpc_listing]
$ python3 run_sie.py solve phi_type_Janus_ring --mode 0 -l 2 -th 0 -a
\end{lstlisting}
\begin{figure}[htbp]
\centering\includegraphics[width=0.95\columnwidth]{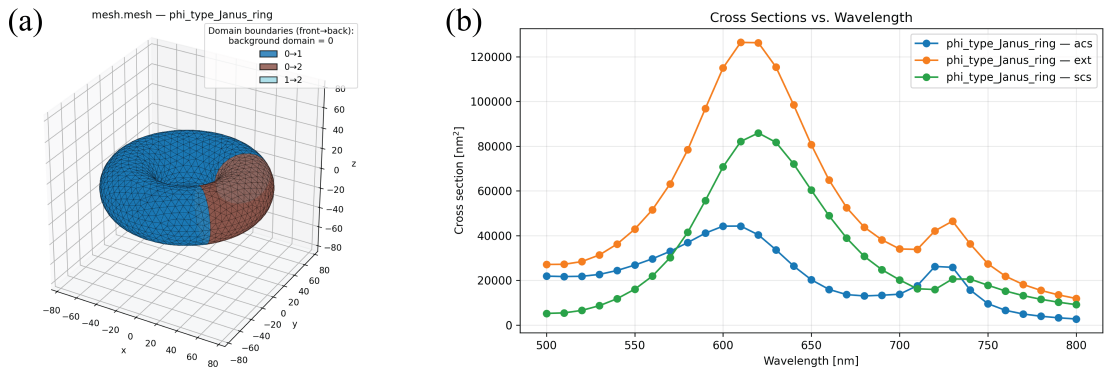}
\caption{Au-cSi Janus ring in free-space. (a) Triangular mesh, and (b) cross-section results for an incident plane wave that propagates towards $+z$ and is $x$-polarized.}
\label{fig:Janus_mesh_cs}
\end{figure}
To examine the optical response of the Janus ring for plane wave illumination, $C_{scs}$, $C_{acs}$ and $C_{ext}$ are visualized in Fig.~\ref{fig:Janus_mesh_cs}(b):
\begin{lstlisting}[style=cpc_listing]
$ python3 pytools/visualization.py phi_type_Janus_ring --root sim_res --plot-csc --subjob 1 --csc scs --csc acs --csc ext --save ./sim_res/phi_type_Janus_ring/out/media/cross_sections.png
\end{lstlisting}

For a detailed inspection of the near-field patterns, a grid of sampling points is generated in the $z = 0$ plane, covering the domain  $x \in [-100,\,100]~\mathrm{nm}$ and $y \in [-105,\,105]~\mathrm{nm}$ with a step size of $0.5~\mathrm{nm}$:
\begin{lstlisting}[style=cpc_listing]
$ python3 run_sie.py make-points phi_type_Janus_ring xy --x -100 100 --y -105 105 --step 0.5 --z0=0 -o points_xy.pos
\end{lstlisting}
Post-processing is then performed at the resonance, $\lambda=620~\mathrm{nm}$:
\begin{lstlisting}[style=cpc_listing]
$ python3 run_sie.py post phi_type_Janus_ring -th 0 -a --lambdas 620 -p points/points_xy.pos
\end{lstlisting}

Finally, two near-field illustrations are presented for different scenarios. More specifically, the total electric field intensity in the $xy$-plane at $\lambda = 620~\mathrm{nm}$ for an incident plane wave and the intensity of its $x$-component for dipole illumination are obtained as shown in Fig.~\ref{fig:Janus_near-fields}:
\begin{lstlisting}[style=cpc_listing]
$ python3 pytools/visualization.py phi_type_Janus_ring --root sim_res --points points_xy.pos --field e --part tot --quantity total --function intensity --mode 2d --subjob 1 --lambda 620 --save ./sim_res/phi_type_Janus_ring/out/media/PW_620nm_Etot_intensity.png
$ python3 pytools/visualization.py phi_type_Janus_ring --root sim_res --points points_xy.pos --field e --part tot --quantity x --function intensity --mode 2d --subjob 3 --log --lambda 620 --save ./sim_res/phi_type_Janus_ring/out/media/dipx_620nm_Ex_intensity.png
\end{lstlisting}
This example highlights the ability of \textls[120]{HELIOS} to simulate complex inhomogeneous particles, while maintaining robust numerical accuracy and providing clear physical insights.

\section{Periodic structures in homogeneous backgrounds}
\label{sec:per_hom}

This section extends the previous formulation to treat periodic structures embedded in homogeneous media. We first outline the necessary theoretical adaptations; specifically, the application of Floquet-periodic boundary conditions and the use of Ewald's transformation for Green's tensor. Since Ewald's summation relies heavily on the complementary error function, the solver utilizes a precomputed binary lookup table (\texttt{erfc.bin}) for acceleration. This file is provided with the source code; however, we also include a dedicated utility to regenerate it locally if any compatibility issues are encountered. Hence, the user can make a new \texttt{erfc.bin} with desired maximum and increment values:
\begin{lstlisting}[style=cpc_listing]
$ make table
$ python run_sie.py make-table erfc.bin --maxRe 10.0 --maxIm 10.0 --incRe 0.1 --incIm 0.1
\end{lstlisting}
Subsequently, we present two numerical examples, i.e.~a photonic crystal and a fishnet structure, to illustrate the handling of periodic lattices and the extraction of scattering parameters. 

\subsection{Theory}
\label{subsec:per_hom_theory}

\begin{figure}[htbp]
\centering\includegraphics[width=0.35\columnwidth]{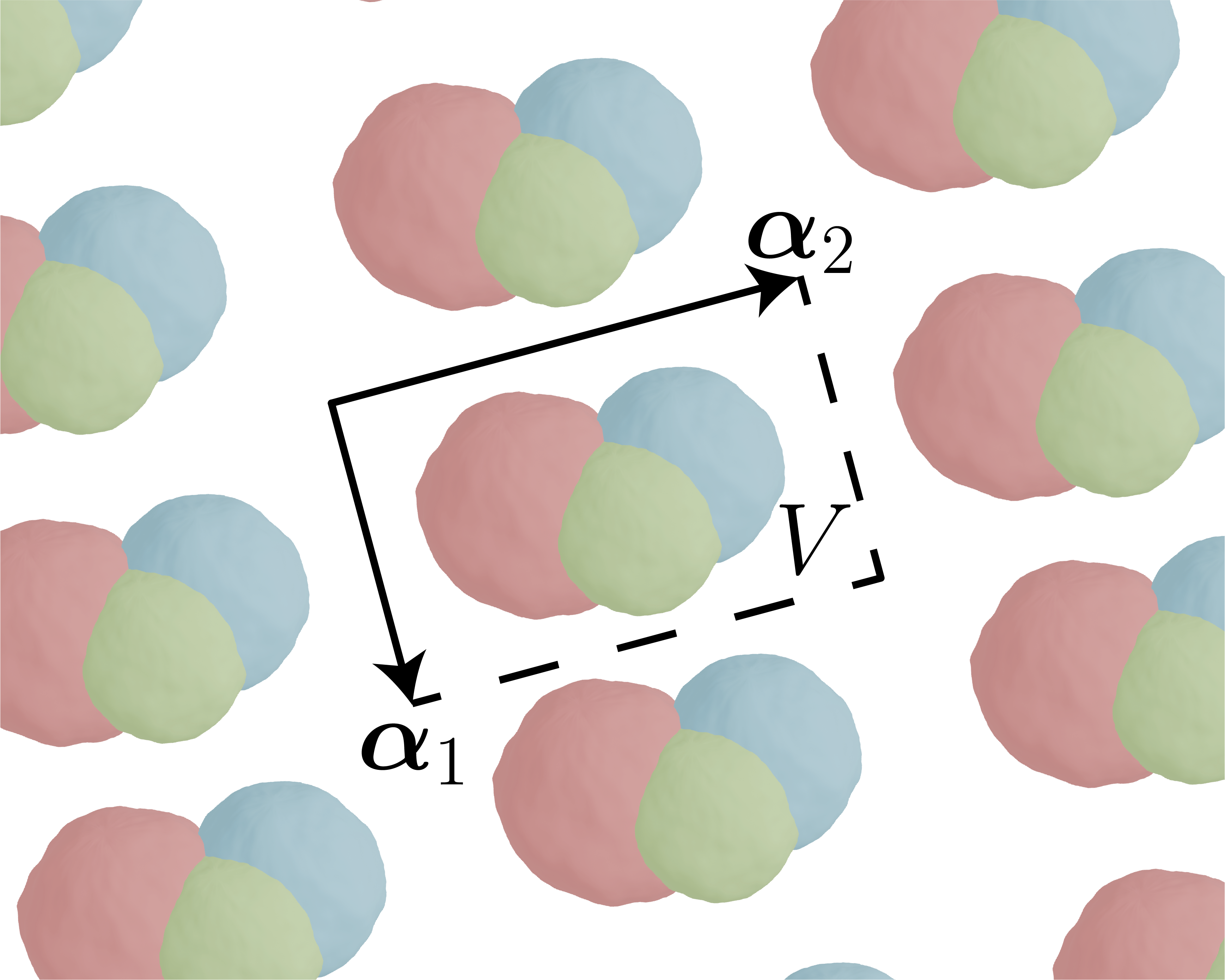}
\caption{Multiple homogeneous scatterers in a unit cell of a periodic structure embedded in a homogeneous background.}
\label{fig:domains_per_hom}
\end{figure}

The modeling of periodic structures relies on the definition of a 2D lattice. In the present framework, the lattice possesses translational symmetry along the $x$ and $y$ axes. A translation vector $\mathbf{t}$ within this lattice is expressed as a linear combination $\mathbf{t} = c_1 \boldsymbol{\alpha}_1 + c_2 \boldsymbol{\alpha}_2$, where $c_1,c_2 \in \mathbb{Z}$ and $\boldsymbol{\alpha}_1,\boldsymbol{\alpha}_2$ denote the primitive lattice vectors. The fundamental repeating unit of the lattice is the unit cell, labeled as $V$ in Fig.~\ref{fig:domains_per_hom}. Additionally, the irreducible representations of the translation group are defined by wavevectors $\mathbf{k}$ situated within the first Brillouin zone \cite{Ludwig_1996}, while the 1D spaces forming the bases of these representations consist of Bloch functions that satisfy the Floquet-periodic boundary conditions
\begin{equation}
    \mathbf{U_k}(\mathbf{r}-\mathbf{t}) = \mathbf{U_k}(\mathbf{r}) \exp(-j \mathbf{k} \cdot \mathbf{t}) \,.
    \label{eq:Floquet_per_BC}
\end{equation}
The projections of the incident electric and magnetic fields onto this space can be written as $\mathbf{E}^{\text{inc}}_{i,\mathbf{k}}$ and $\mathbf{H}^{\text{inc}}_{i,\mathbf{k}}$, respectively. Should the volume current distributions in Eqs.~\eqref{eq:double_curl_Ei} and~\eqref{eq:double_curl_Hi} not be Floquet-periodic, the incident fields $\mathbf{E}^{\text{inc}}_i$ and $\mathbf{H}^{\text{inc}}_i$ may be expressed as linear combinations of Floquet-periodic components $\mathbf{E}^{\text{inc}}_{i,\mathbf{k}}$ and $\mathbf{H}^{\text{inc}}_{i,\mathbf{k}}$.

The evaluation of the EFIE and MFIE is restricted to the boundaries lying within the volume $V$ of the unit cell. Hence, for periodic structures Eqs.~\eqref{eq:EFIE} and~\eqref{eq:MFIE} can be rewritten as
\begin{align}
    j \omega \mu_i \left.\mathbf{\mathcal{D}}_{i,\mathbf{k}}\left\{\mathbf{J}_{s,i,\mathbf{k}} \right\} \left(\mathbf{r}\right) \right|_{\tan} + \left.\mathbf{\mathcal{K}}_{i,\mathbf{k}}\left\{\mathbf{M}_{s,i,\mathbf{k}} \right\}\left(\mathbf{r}\right) \right|_{\tan} &= \left.\mathbf{E}_{i,\mathbf{k}}^{\mathrm{inc}}\left(\mathbf{r}\right) \right|_{\tan} \,, \label{eq:EFIE_unit_cell} \\
    j \omega \varepsilon_i \left.\mathbf{\mathcal{D}}_{i,\mathbf{k}}\left\{\mathbf{M}_{s,i,\mathbf{k}} \right\} \left(\mathbf{r}\right) \right|_{\tan} - \left.\mathbf{\mathcal{K}}_{i,\mathbf{k}}\left\{\mathbf{J}_{s,i,\mathbf{k}} \right\}\left(\mathbf{r}\right) \right|_{\tan} &= \left.\mathbf{H}_{i,\mathbf{k}}^{\mathrm{inc}}\left(\mathbf{r}\right) \right|_{\tan} \,. \label{eq:MFIE_unit_cell}
\end{align}
Each Floquet component of the surface current densities, $\mathbf{J}_{s,i,\mathbf{k}}$ and $\mathbf{M}_{s,i,\mathbf{k}}$, is solved independently. The dyadic pseudo-periodic Green's tensor is defined as
\begin{equation}
    \overline{\mathbf{G}}_{i,\mathbf{k}}(\mathbf{r},\mathbf{r}') =
    \sum_{\mathbf{t}} \exp(j\mathbf{k}\cdot\mathbf{t}) \overline{\mathbf{G}}_{i}(\mathbf{r}-\mathbf{t},\mathbf{r}^{\prime}) \,.
    \label{eq:Green_per_sum}
\end{equation}
Expanding the right-hand side, we obtain
\begin{equation}
    \overline{\mathbf{G}}_i\left(\mathbf{r}, \mathbf{r}^{\prime}\right)=\left(\overline{\mathbf{1}}+\frac{\nabla \nabla}{k_i^{2}}\right) \sum_{\mathbf{t}} \frac{\exp(j k_i \left| \mathbf{R_t} \right| + j \mathbf{k} \cdot \mathbf{t})}{4 \pi \left| \mathbf{R_t} \right|}\,,
    \label{eq:Green_per_full}
\end{equation}
where $\mathbf{R_t} = \mathbf{r} - \mathbf{r}^{\prime} - \mathbf{t}$. Due to the slow convergence of this series, Ewald's transformation \cite{Jordan_1986,Stevanovic_2006} is employed to decompose it into two rapidly convergent sums,
\begin{equation}
    g_{i,\mathbf{k}}\left(\mathbf{r}, \mathbf{r}^{\prime}\right) = g_{i,\mathbf{k}}^{sp}\left(\mathbf{r}, \mathbf{r}^{\prime}\right) + g_{i,\mathbf{k}}^{rp}\left(\mathbf{r}, \mathbf{r}^{\prime}\right)\,,
    \label{eq:Ewald_sum}
\end{equation}
where $g_{i,\mathbf{k}}\left(\mathbf{r}, \mathbf{r}^{\prime}\right)$ represents the sum in Eq.~\eqref{eq:Green_per_full}, $g_{i,\mathbf{k}}^{sp}\left(\mathbf{r}, \mathbf{r}^{\prime}\right)$ is the spatial domain term, and $g_{i,\mathbf{k}}^{rp}\left(\mathbf{r}, \mathbf{r}^{\prime}\right)$ is the reciprocal space term. More specifically, the spatial domain term is given by
\begin{equation}
    g_{i,\mathbf{k}}^{sp}(\mathbf{r},\mathbf{r}') = \frac{1}{8\pi} \sum_{\mathbf{t}} \exp(j\mathbf{k}\cdot\mathbf{t}) \sum_{\pm} \frac{\exp(\pm j k_i |\mathbf{R_t}|)}{|\mathbf{R_t}|}\,\mathrm{erfc} \left( |\mathbf{R_t}|f \pm \frac{j k_i}{2f} \right)\,.
    \label{eq:Ewald_spatial}
\end{equation}
The reciprocal space term involves the reciprocal lattice vectors $\boldsymbol{\beta}_1$ and $\boldsymbol{\beta}_2$, which can be expressed as $\boldsymbol{\beta}_1 = (2\pi / A^2)\boldsymbol{\alpha}_2 \times (\boldsymbol{\alpha}_1 \times \boldsymbol{\alpha}_2)$ and $\boldsymbol{\beta}_2 = (2\pi / A^2) \boldsymbol{\alpha}_1 \times (\boldsymbol{\alpha}_2 \times \boldsymbol{\alpha}_1)$, where $A = \left| \boldsymbol{\alpha}_1 \times \boldsymbol{\alpha}_2 \right|$. By defining $\mathbf{u} = d_1 \boldsymbol{\beta}_1 + d_2 \boldsymbol{\beta}_2$ ($d_1,d_2\in\mathbb{Z}$), one obtains
\begin{equation}
    g_{i,\mathbf{k}}^{rp}(\mathbf{r},\mathbf{r}') = \frac{1}{4 A} \sum_{\mathbf{u}} \exp[j (\mathbf{k}-\mathbf{u}) \cdot (\mathbf{r} - \mathbf{r}^{\prime})] \sum_{\pm} \frac{\exp(\pm \gamma_{i,\mathbf{k},\mathbf{u}} R_\perp)}{\gamma_{i,\mathbf{k},\mathbf{u}}} \, \mathrm{erfc} \left( \frac{\gamma_{i,\mathbf{k},\mathbf{u}}}{2f} \pm R_\perp f \right)\,,
    \label{eq:Ewald_reciprocal}
\end{equation}
where $\gamma_{i,\mathbf{k},\mathbf{u}} = \sqrt{|\mathbf{k}-\mathbf{u}|^2 - k_i^2}$ and $R_\perp = \left(\mathbf{r} -\mathbf{r}^{\prime} \right) \cdot \left( \boldsymbol{\alpha}_1 \times \boldsymbol{\alpha}_2 \right) / A$. The parameter $f$ governs the convergence of both series; an optimal value is typically $f = \pi/\sqrt{A}$ as reported in \cite{Jordan_1986}.

From Eqs.~\eqref{eq:EFIE_unit_cell} and~\eqref{eq:MFIE_unit_cell}, it is evident that applying the PMCHWT formulation yields a linear system analogous to Eq.~\eqref{eq:TPMCHWT}, following the discretization of the unit cell's scatterer(s) with a triangular mesh $\mathcal{M}$ and the application of the Galerkin scheme with RWG functions. A critical implementation detail arises when elements of $\mathcal{M}$ extend to one or more pairs of opposite boundaries of the unit cell. In such cases, translational symmetry must be explicitly enforced to maintain current continuity. Thus, the Floquet-periodic boundary conditions impose
\begin{align}
    \mathbf{J}_{s,i,\mathbf{k}}(\mathbf{r}-\mathbf{t}) &= \exp(-j \mathbf{k} \cdot \mathbf{t}) \mathbf{J}_{s,i,\mathbf{k}}(\mathbf{r})\,, \label{eq:Floquet_Js} \\
    \mathbf{M}_{s,i,\mathbf{k}}(\mathbf{r}-\mathbf{t}) &= \exp(-j \mathbf{k} \cdot \mathbf{t}) \mathbf{M}_{s,i,\mathbf{k}}(\mathbf{r})\,,
    \label{eq:Floquet_Ms}
\end{align}
ensuring consistent flow across periodic boundaries. A detailed analysis is provided in \cite{Gallinet_2010}.

\subsection{Simulating periodic structures in homogeneous backgrounds}
\label{subsec:per_hom_simulation}

In this subsection, we demonstrate the simulation of periodic structures embedded in homogeneous backgrounds (\texttt{----mode 1}). The following examples illustrate the handling of infinite 2D periodic lattices and extraction of spectral quantities, such as reflectance and transmittance. We consider two distinct cases: a dielectric photonic crystal and a fishnet structure. Along with the standard solver arguments introduced previously (namely \texttt{-l}, \texttt{-th}, and \texttt{-a}), the flag \texttt{----etm} is utilized here to specify the number of terms in the evaluation of Green's function with Ewald's method. The only limitation is the code's singular behavior when the wavelength is equal to the period.

\subsubsection{Photonic crystal}
\label{subsubsec:photonic_crystal}

\begin{figure}[H]
\centering\includegraphics[width=0.95\columnwidth]{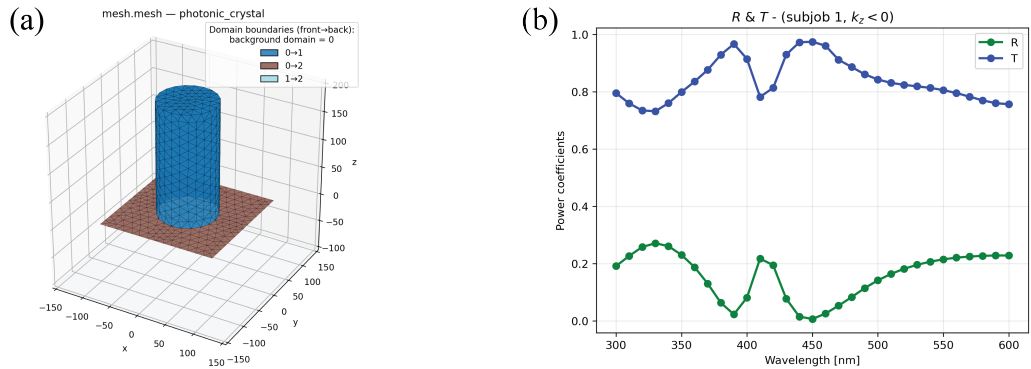}
\caption{Photonic crystal example. (a) Unit cell's triangular mesh (the dimensions of the visualization cell exceed those of the unit cell), and (b) reflectance and transmittance as a function of wavelength for a plane wave illumination that propagates towards $-z$ and is $x$-polarized.}
\label{fig:crystal_mesh_rt}
\end{figure}

To demonstrate the capabilities of \textls[120]{HELIOS} with periodic structures, we first simulate a photonic crystal. The simulation environment is initialized with the \texttt{prepare} command:
\begin{lstlisting}[style=cpc_listing]
$ python3 run_sie.py prepare photonic_crystal --mode 1 --materials eps --spline
\end{lstlisting}
Before launching the solver, the mesh of the unit cell is inspected. The visualization, as shown in Fig.~\ref{fig:crystal_mesh_rt}(a), is saved to the results directory:
\begin{lstlisting}[style=cpc_listing]
$ mkdir sim_res/photonic_crystal/out/media/
$ python3 pytools/plot_mesh.py --sim photonic_crystal --save ./sim_res/photonic_crystal/out/media/mesh.png
\end{lstlisting}
The periodic structure is an infinite square array of pillars with refractive index $n=3.36$, diameter $D=100~\mathrm{nm}$, height $h=200~\mathrm{nm}$, and a substrate of the
same material. Also, the lattice periods are $p_x=p_y=200~\mathrm{nm}$. Proper definition of the extra periodic simulation parameters \texttt{px py cx cy z} in \texttt{config.txt} is essential. Specifically, \texttt{px} and \texttt{py} define the lattice periodicity along the $x$ and $y$ axes, respectively, while \texttt{cx} and \texttt{cy} set the center of the unit cell in the $xy$-plane. The final parameter \texttt{z} establishes the vertical position of the mesh cutting plane below which everything is clipped, here set at $z = -100~\mathrm{nm}$. Consequently, any mesh elements lying on or outside the boundaries defined by these parameters (i.e.~\texttt{px}, \texttt{py}, and \texttt{z}) are clipped. In this example, although the original input mesh includes a finite slab spanning $x,y \in [-100,100]~\mathrm{nm}$ and $z \in [-100,0]~\mathrm{nm}$, the lateral periodic boundaries and the cutoff plane trim the geometry, leaving only the top planar interface included in the computation, as in Fig.~\ref{fig:crystal_mesh_rt}(a). Next, the problem is solved by using \texttt{----mode 1} to indicate a periodic simulation:
\begin{lstlisting}[style=cpc_listing]
$ python3 run_sie.py solve photonic_crystal --mode 1 -l 2 -th 0 -a --etm 4
\end{lstlisting}

A primary quantity of interest for periodic structures is the spectral response, specifically the reflectance ($R$) and transmittance ($T$). Hence, we generate the points file \texttt{rt.pos}, defining the far-field measurement planes within the limits of the unit cell's $xy$-plane ($x,y \in [-100,100]~\mathrm{nm}$ and $z \in \{-1000,1200\}~\mathrm{nm}$), and then compute the desired coefficients using the \texttt{post} command:
\begin{lstlisting}[style=cpc_listing]
$ python3 run_sie.py make-points photonic_crystal xy --x -100 100 --y -100 100 --step 5 --z0=-1000,1200 -o rt.pos
$ python3 run_sie.py post photonic_crystal --mode 1 -p points/rt.pos --etm 4 -a -th 0
\end{lstlisting}
The resulting $R$ and $T$ spectra, for an incident plane wave that propagates towards $-z$ and is $x$-polarized, can be plotted and saved, as displayed in Fig.~\ref{fig:crystal_mesh_rt}(b), with the following command:
\begin{lstlisting}[style=cpc_listing]
$ python pytools/visualization.py photonic_crystal --root sim_res --plot-rt --points rt.pos --subjob 1 --rt-prop auto --save ./sim_res/photonic_crystal/out/media/RTcoeffs.png
\end{lstlisting}

To visualize the near-field distribution, we define a slice in the $y=0$ plane and perform post-processing at $\lambda = 410~\mathrm{nm}$:
\begin{lstlisting}[style=cpc_listing]
$ python3 run_sie.py make-points photonic_crystal xz --x -100 100 --z -100 390 --stepx 2 --stepz 3.5 --y0=0 -o points_xz.pos
$ python3 run_sie.py post photonic_crystal --mode 1 -p points/points_xz.pos --etm 4 -a -th 0 --lambdas 410
\end{lstlisting}
\begin{figure}[H]
\centering\includegraphics[width=0.95\columnwidth]{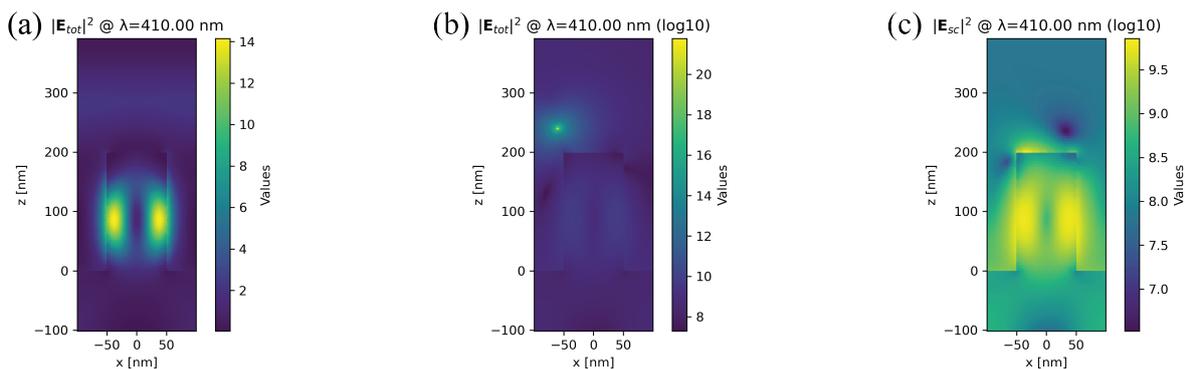}
\caption{Near-field intensity distributions at $\lambda=410~\mathrm{nm}$ for (a) an incident plane wave that propagates towards $-z$ and is $x$-polarized, and (b), (c) an in-plane dipole excitation ($x = -60~\mathrm{nm},\, z = 240~\mathrm{nm},\, \theta = 76^{\circ},\, \varphi = 0^{\circ}$).}
\label{fig:crystal_near_field}
\end{figure}
Finally, the total electric field intensity is visualized for both plane wave (\texttt{----subjob 1}) and dipole excitations (\texttt{----subjob 2}) as depicted in Figs.~\ref{fig:crystal_near_field}(a) and~\ref{fig:crystal_near_field}(b), respectively. For the dipole illumination scenario, the scattered electric field intensity is also shown in Fig.~\ref{fig:crystal_near_field}(c). All these PNG images are produced with the commands:
\begin{lstlisting}[style=cpc_listing]
$ python3 pytools/visualization.py photonic_crystal --root sim_res --points points_xz.pos --field e --part tot --quantity total --function intensity --mode 2d --subjob 1 --lambda 410 --save ./sim_res/photonic_crystal/out/media/PW_410nm_Etot_intensity.png
$ python3 pytools/visualization.py photonic_crystal --root sim_res --points points_xz.pos --field e --part tot --quantity total --function intensity --mode 2d --subjob 2 --log --lambda 410 --save ./sim_res/photonic_crystal/out/media/dip_410nm_Etot_intensity.png
$ python3 pytools/visualization.py photonic_crystal --root sim_res --points points_xz.pos --field e --part sc --quantity total --function intensity --mode 2d --subjob 2 --log --lambda 410 --save ./sim_res/photonic_crystal/out/media/dip_410nm_Esc_intensity.png          
\end{lstlisting}

\subsubsection{Fishnet structure}
\label{subsubsec:fishnet_structure}

\begin{figure}[htbp]
\centering\includegraphics[width=0.95\columnwidth]{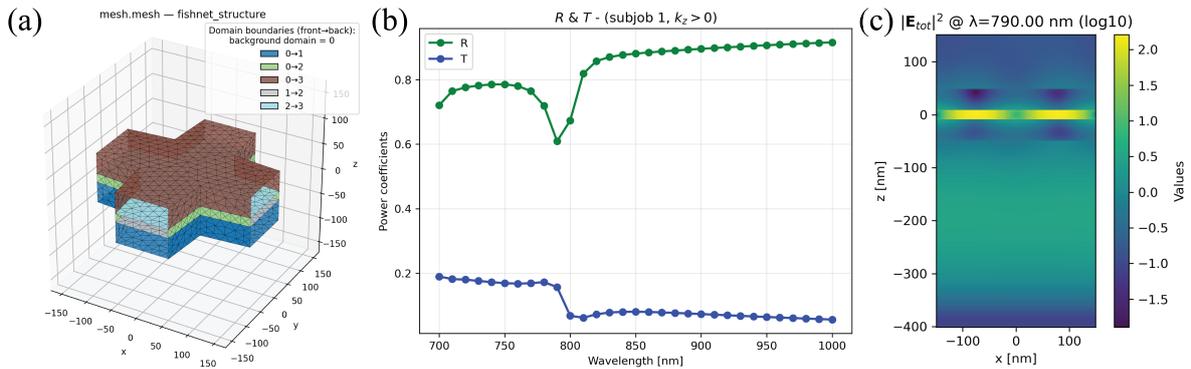}
\caption{Fishnet structure example. (a) Unit cell's triangular mesh, (b) $R$ and $T$ spectra, and (c) total electric near-field intensity at $\lambda=790~\mathrm{nm}$ for a $p$-polarized incident plane wave propagating towards ($\theta = 6^{\circ},\, \varphi = 0^{\circ}$).}
\label{fig:fishnet_res}
\end{figure}

The second example considers a fishnet structure. The workflow is identical to the previous case. We begin by preparing the simulation and verifying the mesh, which is depicted in Fig.~\ref{fig:fishnet_res}(a):
\begin{lstlisting}[style=cpc_listing]
$ python3 run_sie.py prepare fishnet_structure --mode 1 --materials eps --spline
$ mkdir sim_res/fishnet_structure/out/media/
$ python3 pytools/plot_mesh.py --sim fishnet_structure --save ./sim_res/fishnet_structure/out/media/mesh.png
\end{lstlisting}
All the details of the system can be found in \cite{Gallinet_2010}. Notably, the resulting geometry is a continuous fishnet structure that extends over the entire space, since the scatterers occupy the entire unit cell. The solver is then executed using the periodic mode:
\begin{lstlisting}[style=cpc_listing]
$ python3 run_sie.py solve fishnet_structure --mode 1 -l 2 -th 0 -a --etm 4
\end{lstlisting}
We subsequently compute the $R$ and $T$ coefficients to identify spectral features. The system is excited with a $p$-polarized plane wave that propagates towards the direction ($\theta = 6^{\circ},\, \varphi = 0^{\circ}$). Results are visualized in Fig.~\ref{fig:fishnet_res}(b):
\begin{lstlisting}[style=cpc_listing]
$ python3 run_sie.py make-points fishnet_structure xy --x -150 150 --y -150 150 --step 5 --z0=-1000,1000 -o rt.pos
$ python3 run_sie.py post fishnet_structure --mode 1 -p points/rt.pos --etm 4 -a -th 0
$ python3 pytools/visualization.py fishnet_structure --root sim_res --plot-rt --points rt.pos --subjob 1 --rt-prop auto --save ./sim_res/fishnet_structure/out/media/RTcoeffs.png
\end{lstlisting}

A near-field map is generated at $\lambda = 790~\mathrm{nm}$. We define the sampling grid in the $y=0$ plane with a step of 2.5 nm and run the post-processing:
\begin{lstlisting}[style=cpc_listing]
$ python3 run_sie.py make-points fishnet_structure xz --x -150 150 --z -400 150 --step 2.5 --y0=0 -o points_xz.pos
$ python3 run_sie.py post fishnet_structure --mode 1 -p points/points_xz.pos --etm 4 -a -th 0 --lambdas 790
\end{lstlisting}
Finally, the total electric field intensity distribution for the selected plane wave illumination is visualized, as in Fig.~\ref{fig:fishnet_res}(c):
\begin{lstlisting}[style=cpc_listing]
$ python3 pytools/visualization.py fishnet_structure --root sim_res --points points_xz.pos --field e --part tot --quantity total --function intensity --mode 2d --subjob 1 --log --lambda 790 --save ./sim_res/fishnet_structure/out/media/PW_790nm_Etot_intensity.png  
\end{lstlisting}

\section{Isolated scatterers in layered backgrounds}
\label{sec:iso_layered}

In this section, we address the modeling of isolated scatterers embedded in stratified media. This configuration corresponds to \texttt{----mode 2} in \textls[120]{HELIOS}. We first derive the necessary Green's tensor components for layered environments, including the evaluation of Sommerfeld integrals. Subsequently, we provide three examples, i.e.~a sphere on a substrate, a hole in a metallic film, and a hole in a multilayer stack, to demonstrate the solver's capabilities.

\subsection{Theory}
\label{subsec:iso_layered_theory}

As demonstrated in the preceding sections, the cornerstone of the SIE method is the dyadic Green's tensor. For stratified media, this tensor must incorporate the reflection and transmission effects occurring at each background interface.
\begin{figure}[htbp]
\centering\includegraphics[width=0.95\columnwidth]{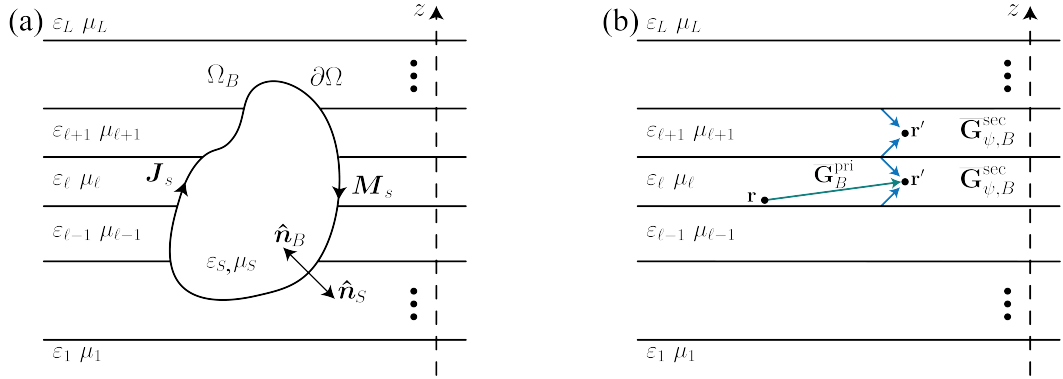}
\caption{Homogeneous scatterer in a stratified background. (a) Geometry for the scattering problem analysis, and (b) contributions of primary (green) and secondary (blue) dyadic Green's tensor terms in layered media.}
\label{fig:domains_layered}
\end{figure}
Let us consider the system depicted in Fig.~\ref{fig:domains_layered}(a), where a homogeneous scatterer (domain $\Omega_S$) is embedded in a layered background (domain $\Omega_B$). The dyadic Green's tensor describing $\Omega_B$ is expressed as \cite{Chen_DGLM}
\begin{equation}
    \overline{\mathbf{G}}_{\psi,B}(\mathbf{r},\mathbf{r}^{\prime}) = \overline{\mathbf{G}}^{\mathrm{pri}}_{B}(\mathbf{r},\mathbf{r}^{\prime}) + \overline{\mathbf{G}}^{\mathrm{sec}}_{\psi,B}(\mathbf{r},\mathbf{r}^{\prime}) \,,
    \label{eq:dyadic_G_layered}
\end{equation}
where $\overline{\mathbf{G}}^{\mathrm{pri}}_B(\mathbf{r},\mathbf{r}^{\prime})$ represents the primary contribution, $\overline{\mathbf{G}}^{\mathrm{sec}}_{\psi,B}(\mathbf{r},\mathbf{r}^{\prime})$ denotes the secondary contribution arising from the layered medium, and the subscript $\psi$ indicates the field type ($\psi=E$ for electric and $\psi=H$ for magnetic). The primary term corresponds to the dyadic Green's tensor of a homogeneous medium, whereas the secondary term accounts for all scattering interactions between layers and is essential for satisfying the boundary conditions of Maxwell's equations at the interfaces. In the following analysis, the layer indices of the source and observation points are denoted by $m$ and $n$, respectively. We focus on the electric-type scenario ($\psi=E$), noting that the magnetic-type case can be retrieved via the duality principle~\cite[chap.~7]{Chew_1995}. Hence, the primary term is given by
\begin{equation}
    \overline{\mathbf{G}}^{\mathrm{pri}}_B\left(\mathbf{r}, \mathbf{r}^{\prime}\right) = \overline{\mathbf{G}}^{\mathrm{pri}}_{E,B}(\mathbf{r},\mathbf{r}^{\prime}) = \overline{\mathbf{G}}^{\mathrm{pri}}_{H,B}(\mathbf{r},\mathbf{r}^{\prime}) = \left(\overline{\mathbf{1}}+\frac{\nabla \nabla}{k_m^{2}}\right)\frac{\exp(j k_m \left|\mathbf{r}-\mathbf{r^{\prime}}\right|)}{4\pi \left|\mathbf{r}-\mathbf{r^{\prime}}\right|} \,,
    \label{eq:dyadic_G_primary}
\end{equation} 
where $m=n$, since this term contributes exclusively to interactions within the same layer, as shown in Fig.~\ref{fig:domains_layered}(b). The secondary term admits a decomposition into transverse electric ($\mathrm{TE}$) and transverse magnetic ($\mathrm{TM}$) wave contributions \cite{Chew_matrix_friendly}, which reads as
\begin{equation}
    \overline{\mathbf{G}}^{\mathrm{sec}}_{E,B}\left(\mathbf{r}, \mathbf{r}^{\prime}\right) =  \overline{\mathbf{G}}^\mathrm{TE}_{E,B}\left(\mathbf{r}, \mathbf{r}^{\prime}\right) + \frac{1}{k_{nm}^2} \overline{\mathbf{G}}^\mathrm{TM}_{E,B}\left(\mathbf{r}, \mathbf{r}^{\prime}\right) \,,
    \label{eq:dyadic_G_secondary}
\end{equation}
where $k_{nm}^2 = \omega^2 \varepsilon_n \mu_m$ and
\begin{align}
    \overline{\mathbf{G}}^{\mathrm{TE}}_{E,B}\left(\mathbf{r}, \mathbf{r}^{\prime}\right) & =(\nabla \times \hat{\mathbf{z}})\left(\nabla^{\prime} \times \hat{\mathbf{z}}\right) g_B^{\mathrm{TE}}\left(\mathbf{r}, \mathbf{r}^{\prime}\right)\,, 
    \label{eq:dyadic_G_TE} \\
    \overline{\mathbf{G}}^{\mathrm{TM}}_{E,B}\left(\mathbf{r}, \mathbf{r}^{\prime}\right) & =(\nabla \times \nabla \times \hat{\mathbf{z}})\left(\nabla^{\prime} \times \nabla^{\prime} \times \hat{\mathbf{z}}\right) g_B^{\mathrm{TM}}\left(\mathbf{r}, \mathbf{r}^{\prime}\right) \,.
    \label{eq:dyadic_G_TM}
\end{align}
The scalar Green's functions that appear in Eqs.~\eqref{eq:dyadic_G_TE} and~\eqref{eq:dyadic_G_TM} are written in compact form as \cite{Chew_matrix_friendly}
\begin{equation}
    g_B^{\nu}\left(\mathbf{r}, \mathbf{r}^{\prime}\right) = \frac{j}{4 \pi} \int_0^{+\infty} \frac{d k_\rho}{k_{m z} k_\rho} J_0\left(k_\rho \rho\right) F_B^{\nu}\left(k_\rho, z, z^{\prime}\right)\,,
    \label{eq:scalar_g_TE_TM}
\end{equation}
where $\nu \in \{\mathrm{TE}, \mathrm{TM}\}$, $k_\rho$ is the radial wavenumber, $k_{mz} = \sqrt{k_m^2 - k_\rho^2}$, $\rho$ is the radial distance between $\mathbf{r}$ and $\mathbf{r}^{\prime}$ and $F_B^{\nu}(k_\rho, z, z^{\prime})$ is the propagation factor~\cite[chap.~2]{Chew_1995}. The latter determines the reflected or transmitted fields for primary TE or TM fields and is computed using the transfer matrix method~\cite[chap.~8]{Hohenester_book}.

The analytical structure of the secondary Green's tensor, as shown in Eqs.~\eqref{eq:dyadic_G_secondary} to \eqref{eq:dyadic_G_TM}, allows for the transfer of differential operators from the scalar Green's function to RWG test and basis functions. This property is crucial, since curl and divergence operations appear inside surface integrals. However, the existence of $k_\rho$ in the denominator of the integrand in Eq.~\eqref{eq:scalar_g_TE_TM} leads to a singularity as $k_\rho \rightarrow 0$, which is handled as in~\cite[chap.~6]{Chew_2008}. Regarding the singularities of the primary term, they are treated using the same semi-analytical approach employed for homogeneous backgrounds, as mentioned in Section~\ref{subsec:iso_hom_theory}.
\begin{figure}[htbp]
\centering\includegraphics[width=0.95\columnwidth]{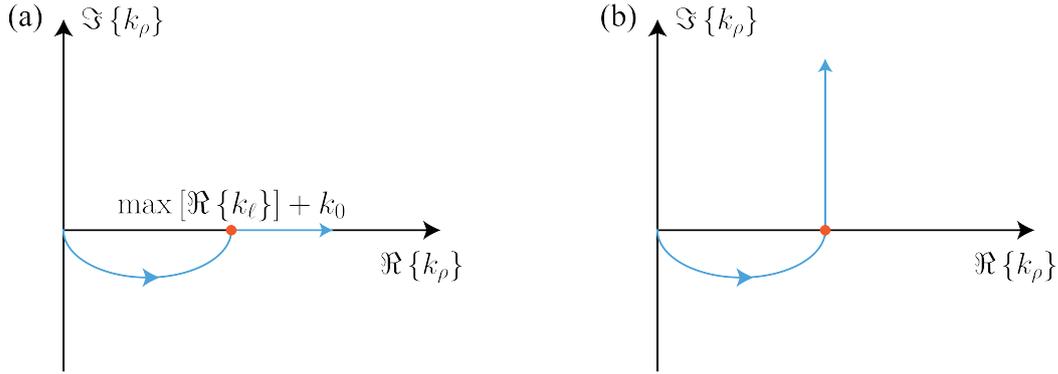}
\caption{Integration paths for Sommerfeld integrals. The first part of the integration follows the semi-ellipse from the origin to the highlighted point of the real axis with value $\max \left[ \Re \left\{ k_{\ell} \right\} \right] + k_0$, where $\max \left[ \Re \left\{ k_{\ell} \right\} \right]$ is the maximum of the real parts of the wavenumbers in the layers of the stratified medium ($\ell = 1,...,L$) and $k_0$ is the free-space wavenumber, which is used as a safety margin. (a) For large $\left| z - z^{\prime} \right|$ values the contour continues along the real axis $\Re\{k_\rho\}$, while (b) for smaller values of $\left| z - z^{\prime} \right|$ the path follows the parallel to the imaginary axis $\Im\{k_\rho\}$.}
\label{fig:integration_path}
\end{figure}

The evaluation of layered medium Green's function requires the numerical computation of Sommerfeld integrals, as depicted in Eq.~\eqref{eq:scalar_g_TE_TM}. For this purpose, we adopt the integration path proposed in \cite{Paulus_2000}, which adapts to the vertical distance $\left| z - z^{\prime} \right|$ between source and observation points. As shown in Fig.~\ref{fig:integration_path}(a), for sufficiently large $\left| z - z^{\prime} \right|$, the path consists of a semi-elliptical contour followed by the remainder of the real axis. Conversely, for small values of $\left| z - z^{\prime} \right|$, the contour continues along a parallel to the imaginary axis, as illustrated in Fig.~\ref{fig:integration_path}(b). This selection ensures an accelerated convergence of the required Sommerfeld integrals.

The final linear system is derived via the implementation of the matrix-friendly formulation \cite{Chen_DGLM, Chew_matrix_friendly, Chew_2008, Chen_2012}, inserting the Green's tensor decomposition from Eq.~\eqref{eq:dyadic_G_secondary} into the PMCHWT formulation. Wherever feasible, derivative operators acting on Green's functions are shifted onto the test and basis RWG functions via partial integration. This results in expressions that are significantly more amenable to numerical evaluation. The final linear system for multiple penetrable scattering bodies in stratified backgrounds is written as 
\begin{equation}
    \begin{split}
    \left(
    \left[
    \begin{array}{cc}
    \mathbf{D}_{B}^{\mathrm{pri}} + \mathbf{D}_{E,B}^{\mathrm{sec}} & 
    \mathbf{K}_{B}^{\mathrm{pri}} + \mathbf{K}_{E,B}^{\mathrm{sec}} \\
    -\mathbf{K}_{B}^{\mathrm{pri}} - \mathbf{K}_{H,B}^{\mathrm{sec}} & 
    \mathbf{D}_{B}^{\mathrm{pri}} + \mathbf{D}_{H,B}^{\mathrm{sec}}
    \end{array}
    \right] + \sum_{i,\, i\neq B}
    \left[
    \begin{array}{cc}
    j \omega \mu_i \mathbf{D}_i & \mathbf{K}_i \\
    -\mathbf{K}_i & j \omega \varepsilon_i \mathbf{D}_i
    \end{array}
    \right]
    \right) \cdot
    \left[
    \begin{array}{c}
    \boldsymbol{\alpha} \\
    \boldsymbol{\beta}
    \end{array}
    \right]
    = \sum_i
    \left[
    \begin{array}{l}
    \mathbf{q}_i^E \\
    \mathbf{q}_i^H
    \end{array}
    \right] \,,
    \end{split}
    \label{eq:TPMCHWT_layered}
\end{equation}
where the first matrix represents the stratified background domain $\Omega_B$ and the summation covers the homogeneous domain(s) of the scatterer(s). The main difference with Eq.~\eqref{eq:TPMCHWT} is that the matrix elements for the background domain differ from the homogeneous case and are expressed as \cite{Chen_DGLM, Chew_matrix_friendly, Chew_2008, Chen_2012}
\begin{align}
    (D_B^{\mathrm{pri}})_{\xi_1\xi_2} &= {}j \omega \mu_{m} \int_{\partial \Omega_B} \mathrm{~d} S \mathbf{f}_{\xi_1,B}(\mathbf{r}) \, \cdot \int_{\partial \Omega_B} \mathrm{~d} S^{\prime} \overline{\mathbf{G}}_B^{\mathrm{pri}}\left(\mathbf{r}, \mathbf{r}^{\prime}\right) \cdot \mathbf{f}_{\xi_2,B}\left(\mathbf{r}^{\prime}\right) \label{eq:Dmn_pri} \,, \\
    (K_B^{\mathrm{pri}})_{\xi_1\xi_2} &= {}\int_{\partial \Omega_B} \mathrm{~d} S \mathbf{f}_{\xi_1,B}(\mathbf{r}) \, \cdot \int_{\partial \Omega_B} \mathrm{~d} S^{\prime}\left[\nabla^{\prime} \times \overline{\mathbf{G}}_B^{\mathrm{pri}}\left(\mathbf{r}, \mathbf{r}^{\prime}\right)\right] \cdot \mathbf{f}_{\xi_2,B}\left(\mathbf{r}^{\prime}\right) \label{eq:Kmn_pri} \,, \\
    (D_{E,B}^{\mathrm{sec}})_{\xi_1\xi_2} &= {}j \omega \mu_{m}
    \left\{
    \left \langle \nabla \cdot \mathbf{f}_{\xi_1,B}(\mathbf{r}) \left |\, k_{n m}^{-2} \partial_z \partial_{z^{\prime}} g_B^{\mathrm{TM}}\left(\mathbf{r}, \mathbf{r}^{\prime}\right)-g_B^{\mathrm{TE}}\left(\mathbf{r}, \mathbf{r}^{\prime}\right) \,\right| \nabla^{\prime} \cdot \mathbf{f}_{\xi_2,B} \left(\mathbf{r}^{\prime}\right) \right \rangle \right. \nonumber \\
    &\hspace{1cm}- \left \langle \hat{\mathbf{z}} \cdot \mathbf{f}_{\xi_1,B}(\mathbf{r}) \left|\, \mu_n \mu_m^{-1} \partial_{z^{\prime}} g_B^{\mathrm{TM}}\left(\mathbf{r}, \mathbf{r}^{\prime}\right) + \partial_z g_B^{\mathrm{TE}}\left(\mathbf{r}, \mathbf{r}^{\prime}\right) \,\right| \nabla^{\prime} \cdot \mathbf{f}_{\xi_2,B}\left(\mathbf{r}^{\prime}\right) \right \rangle \nonumber \\
    &\hspace{1cm}- \left \langle \nabla \cdot \mathbf{f}_{\xi_1,B}(\mathbf{r}) \left|\, \varepsilon_m \varepsilon_n^{-1} \partial_z g_B^{\mathrm{TM}}\left(\mathbf{r}, \mathbf{r}^{\prime}\right) + \partial_{z^{\prime}} g_B^{\mathrm{TE}}\left(\mathbf{r}, \mathbf{r}^{\prime}\right) \,\right| \hat{\mathbf{z}} \cdot \mathbf{f}_{\xi_2,B}\left(\mathbf{r}^{\prime}\right) \right \rangle \nonumber \\
    &\hspace{1cm}+ \left \langle \hat{\mathbf{z}} \cdot \mathbf{f}_{\xi_1,B}(\mathbf{r}) \left|\, k_{m n}^2 g_B^{\mathrm{TM}}\left(\mathbf{r}, \mathbf{r}^{\prime}\right)-\partial_z \partial_{z^{\prime}} g_B^{\mathrm{TE}}\left(\mathbf{r}, \mathbf{r}^{\prime}\right) \,\right| \hat{\mathbf{z}} \cdot \mathbf{f}_{\xi_2,B}\left(\mathbf{r}^{\prime}\right) \right \rangle \nonumber \\
    &\hspace{1cm}+ \left \langle \hat{\mathbf{x}} \cdot \mathbf{f}_{\xi_1,B}(\mathbf{r}) \left|\, g_{B,s}^{\mathrm{TE}}\left(\mathbf{r}, \mathbf{r}^{\prime}\right) \,\right| \hat{\mathbf{x}} \cdot \mathbf{f}_{\xi_2,B}\left(\mathbf{r}^{\prime}\right) \right \rangle  \nonumber \\
    & \left.\hspace{1cm}+ \left \langle \hat{\mathbf{y}} \cdot \mathbf{f}_{\xi_1,B}(\mathbf{r}) \left|\, g_{B,s}^{\mathrm{TE}}\left(\mathbf{r}, \mathbf{r}^{\prime}\right) \,\right| \hat{\mathbf{y}} \cdot \mathbf{f}_{\xi_2,B}\left(\mathbf{r}^{\prime}\right) \right \rangle
    \right\}  \label{eq:Dmn_sec} \,, \\
    (K_{E,B}^{sec})_{\xi_1\xi_2} &= {}
    \left \langle \hat{\mathbf{z}} \cdot \mathbf{f}_{\xi_1,B}(\mathbf{r}) \left|\, k_{mn}^2 \partial_{\rho} g_B^{\mathrm{TM}}\left(\mathbf{r}, \mathbf{r}^{\prime}\right) \,\right| \hat{\mathbf{z}} \cdot\left[\mathbf{f}_{\xi_2,B}(\mathbf{r}^{\prime}) \times\left(\mathbf{r}-\mathbf{r}^{\prime}\right)\right] \rho^{-1} \right \rangle \nonumber \\
    &-\hspace{0.1cm} \left \langle \hat{\mathbf{z}} \cdot\left[\mathbf{f}_{\xi_1,B}(\mathbf{r}) \times\left(\mathbf{r}-\mathbf{r}^{\prime}\right)\right] \rho^{-1} \left |\, k_m^2 \partial_{\rho} g_B^{\mathrm{TE}}\left(\mathbf{r}, \mathbf{r}^{\prime}\right) \,\right| \hat{\mathbf{z}} \cdot \mathbf{f}_{\xi_2,B} \left(\mathbf{r}^{\prime}\right) \right \rangle \nonumber \\
    &+\hspace{0.1cm} \left \langle \hat{\mathbf{z}} \cdot\left[\mathbf{f}_{\xi_1,B}(\mathbf{r}) \times\left(\mathbf{r}-\mathbf{r}^{\prime}\right)\right] \rho^{-1} \left|\, \partial_{\rho} \partial_{z^\prime} g_B^{\mathrm{TE}}\left(\mathbf{r}, \mathbf{r}^{\prime}\right) \,\right| \nabla^{\prime} \cdot \mathbf{f}_{\xi_2,B}\left(\mathbf{r}^{\prime}\right) \right \rangle \nonumber \\
    &-\hspace{0.1cm} \left \langle \nabla \cdot \mathbf{f}_{\xi_1,B}(\mathbf{r}) \left|\, \varepsilon_m \varepsilon_n^{-1} \partial_{\rho} \partial_{z} g_B^{\mathrm{TM}}\left(\mathbf{r}, \mathbf{r}^{\prime}\right) \,\right| \hat{\mathbf{z}} \cdot\left[\mathbf{f}_{\xi_2,B}(\mathbf{r}^{\prime}) \times\left(\mathbf{r}-\mathbf{r}^{\prime}\right)\right] \rho^{-1} \right \rangle
    \label{eq:Kmn_sec} \,,
\end{align}
where the following simplified notation is used
\begin{equation}
    \left \langle \Xi(\mathbf{r}) \left|\, \Phi(\mathbf{r}, \mathbf{r}^{\prime}) \,\right| \Psi(\mathbf{r}^{\prime}) \right \rangle = \int_{\partial \Omega_B}  \int_{\partial \Omega_B} \mathrm{~d} S \mathrm{~d} S^{\prime}\, \Xi(\mathbf{r}) \Phi(\mathbf{r}, \mathbf{r}^{\prime}) \Psi(\mathbf{r}^{\prime}) \,.
\end{equation}

\begin{figure}[htbp]
\centering\includegraphics[width=0.6\columnwidth]{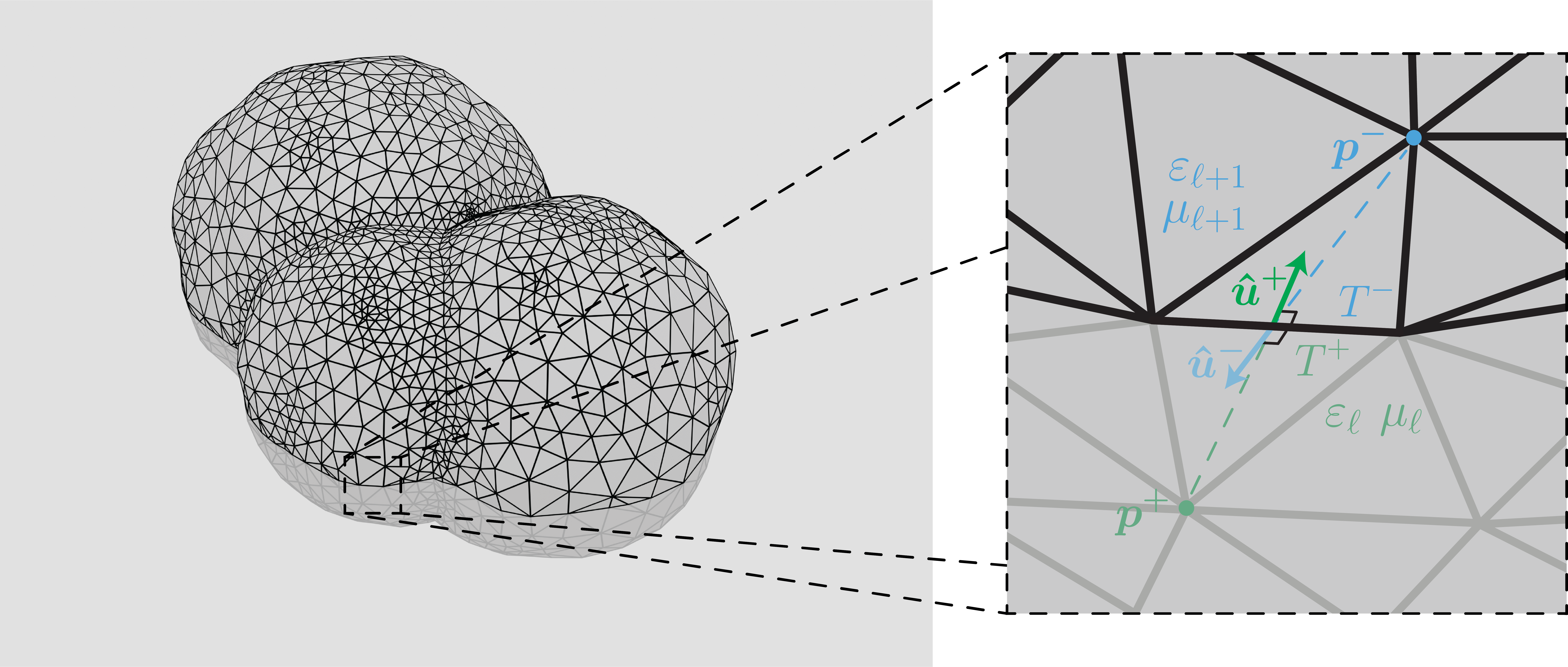}
\caption{Close-up view of a geometric configuration relevant to the line-integral terms. The scatterers go through an interface between two layers with properties ($\varepsilon_\ell, \mu_\ell$) and ($\varepsilon_{\ell + 1}, \mu_{\ell + 1}$). All triangle pairs that are modeling this feature have their common edge exactly on the interface.}
\label{fig:line_integrals}
\end{figure}

During the matrix-filling process, if the surface of the scattering body passes through one or more interfaces, as shown in Fig.~\ref{fig:line_integrals}, then the evaluation of Eq.~\eqref{eq:Kmn_sec} requires the addition of testing ($\ell_t$) and basis ($\ell_b$) line integrals
\begin{align}
    \ell_t &= \varepsilon_m \varepsilon_n^{-1} \int_{\partial T_{\xi_1}} \mathrm{~d} \ell \, \mathbf{f}_{\xi_1,B}(\mathbf{r}) \cdot \hat{\mathbf{u}}_{\xi_1}^\pm \int_{T_{\xi_2}^\pm} \mathrm{~d} S^{\prime}\, \hat{\mathbf{z}} \cdot\left[\mathbf{f}_{\xi_2,B}\left(\mathbf{r}^{\prime}\right) \times\left(\mathbf{r}-\mathbf{r}^{\prime}\right)\right] \rho^{-1} \partial_{\rho} \partial_{z} g_B^{\mathrm{TM}}\left(\mathbf{r}, \mathbf{r}^{\prime}\right)\,, \label{eq:testing_line_int} \\
    \ell_b &= -\int_{\partial T_{\xi_2}} \mathrm{~d} \ell^{\prime}\, \mathbf{f}_{\xi_2,B}\left(\mathbf{r}^{\prime}\right) \cdot \hat{\mathbf{u}}_{\xi_2}^\pm \int_{T_{\xi_1}^\pm} \mathrm{~d} S\, \hat{\mathbf{z}} \cdot\left[\mathbf{f}_{\xi_1,B}(\mathbf{r}) \times\left(\mathbf{r}-\mathbf{r}^{\prime}\right)\right] \rho^{-1} \partial_{\rho} \partial_{z^\prime} g_B^{\mathrm{TE}}\left(\mathbf{r}, \mathbf{r}^{\prime}\right)\,.
    \label{eq:basis_line_int}
\end{align}

In these line integral terms, $\partial T_{\xi_1}$ ($\partial T_{\xi_2}$) denotes the common edge of the triangle pair supporting the RWG test (basis) function $\mathbf{f}_{\xi_1,B}$ ($\mathbf{f}_{\xi_2,B}$), and $\hat{\mathbf{u}}_{\xi_1}^\pm$ ($\hat{\mathbf{u}}_{\xi_2}^\pm$) is the corresponding outward unit normal on that edge. The $\pm$ sign depends on which of the two triangles is being processed by the code at a given moment. The inner surface integral is performed over $T_{\xi_2}^\pm$ ($T_{\xi_1}^\pm$), i.e.~one of the two triangles supporting the basis (test) RWG function $\mathbf{f}_{\xi_2,B}$ ($\mathbf{f}_{\xi_1,B}$). This formulation is particularly convenient as the implementation evaluates matrix entries via half-RWG interactions. The testing line integral is needed only when the testing RWG function straddles the interface and the corresponding RWG or half-RWG basis functions (on either positive or negative triangles) lie within the same two adjacent layers \cite{Chen_DGLM}. An analogous argument determines the requirement of the basis line integral. Notably, ``straddling'' implies that the shared edge of a triangle pair lies exactly on the interface. Finally, the corresponding magnetic-type quantities $\mathbf{D}_{H,B}^{\mathrm{sec}}$ and $\mathbf{K}_{H,B}^{\mathrm{sec}}$ are obtained via the duality principle \cite{Chew_1995}, implemented by the interchanges $(\dots)_E \leftrightarrow {(\dots)_H}$, $\varepsilon \leftrightarrow \mu$ and $(\dots)^{\mathrm{TE}} \leftrightarrow (\dots)^{\mathrm{TM}}$.

A significant numerical challenge in evaluating the layered medium Green's function and its derivatives, that appear in Eqs.~\eqref{eq:Dmn_sec} and~\eqref{eq:Kmn_sec}, arises from the integrands of the underlying Sommerfeld representations. For many configurations, the dominant contribution to these integrals originates from large transverse wavenumbers $k_\rho$, corresponding to the quasi-static regime \cite{Hohenester_nanobem}. This behavior compromises the efficiency of direct numerical quadrature and interpolation-based acceleration strategies, due to the strong and non-smooth dependence on both the radial separation $\rho$ and the vertical coordinates of source and observation points, $z$ and $z^\prime$ respectively \cite{Hohenester_nanobem}. Following the approach of \cite{Hohenester_nanobem, Pratama_2015}, the integrands are decomposed as $g_B^{\nu}\left(\mathbf{r}, \mathbf{r}^{\prime}\right) = g_B^{\nu,sm}\left(\mathbf{r}, \mathbf{r}^{\prime}\right) + g_B^{\nu,qs}\left(\mathbf{r}, \mathbf{r}^{\prime}\right)$, where $g_B^{\nu,qs}\left(\mathbf{r}, \mathbf{r}^{\prime}\right)$ represents the quasi-static part and $g_B^{\nu,sm}\left(\mathbf{r}, \mathbf{r}^{\prime}\right)$ the smooth part. The latter is a numerically well behaved quantity and its evaluation is accurate; however, to further accelerate repeated Green's function calculations, $g_B^{\nu,sm}\left(\mathbf{r}, \mathbf{r}^{\prime}\right)$ and its required derivatives are tabulated on structured grids, which are adapted to the geometry of the problem. When both source and observation points lie in the uppermost or lowermost layer, a 2D grid $g_B^{\nu,sm}(\rho,z+z^\prime)$ suffices. If the points are located within the same interior layer, the use of the two independent 2D grids $g_{B,+}^{\nu,sm}(\rho,z+z^\prime)$ and $g_{B,-}^{\nu,sm}(\rho,z-z^\prime)$ is required. For other configurations, in which the points reside in different layers, a full 3D grid $g_B^{\nu,sm}(\rho,\, z,\, z^\prime)$ is used. This tabulation-interpolation scheme accelerates the computational process, while retaining the required accuracy for surface integral calculations. Hence, it ensures that interpolated values remain stable in parameter regions strongly influenced by the quasi-static contributions, which are calculated analytically and then added back to obtain the full result. Finally, all small distance singularities -- which are isolated in the quasi-static part -- are handled with singularity cancellation techniques \cite{Taylor_2003, DElia_2011, Sarraf_2014}.

\subsection{Simulating nanostructures in stratified backgrounds}
\label{subsec:iso_layered_simulation}

This section details the simulation of isolated scatterers within stratified media, enabled by the \texttt{----mode 2} flag. These examples demonstrate the solver's ability to account for substrate effects and complex multilayer environments using the formulation described in the previous section.

\subsubsection{Silver sphere on SiO\textsubscript{2}}
\label{subsubsec:Ag_sph_SiO2}

\begin{figure}[htbp]
\centering\includegraphics[width=0.95\columnwidth]{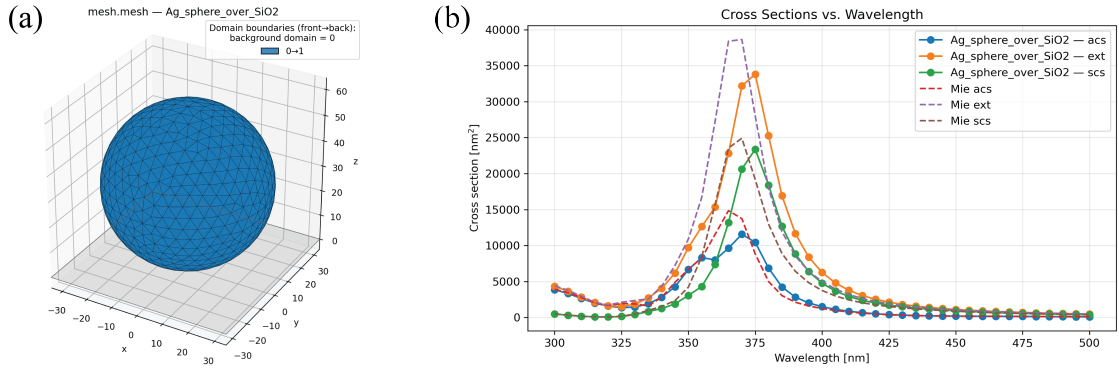}
\caption{Ag sphere ($R=30~\mathrm{nm}$) on a SiO\textsubscript{2} substrate. (a) Triangular mesh with the air-SiO\textsubscript{2} interface, and (b) cross-section results for an incident plane wave that propagates towards $-z$ and is $x$-polarized, including a comparison with Mie theory for the same sphere in free-space.}
\label{fig:Ag_sph_mesh_cs}
\end{figure}

We begin with the example of a plasmonic nanoparticle on a dielectric substrate; more specifically, an Ag sphere of radius $R=30~\mathrm{nm}$ located on top of a SiO\textsubscript{2} half-space. This configuration allows us to validate the solver's handling of layered media Green's tensor. The simulation environment is initialized and the mesh is saved for inspection, as shown in Fig.~\ref{fig:Ag_sph_mesh_cs}(a):
\begin{lstlisting}[style=cpc_listing]
$ python run_sie.py prepare Ag_sphere_over_SiO2 --mode 2 --materials eps --spline
$ mkdir sim_res/Ag_sphere_over_SiO2/out/media/
$ python3 pytools/plot_mesh.py --sim Ag_sphere_over_SiO2 --save ./sim_res/Ag_sphere_over_SiO2/out/media/mesh.png
\end{lstlisting}
Then, the system is assembled and solved using LU decomposition:
\begin{lstlisting}[style=cpc_listing]
$ python3 run_sie.py solve Ag_sphere_over_SiO2 --mode 2 -l 2 -th 0 -a
\end{lstlisting}

Scattering, absorption, and extinction cross-sections are calculated and plotted. To illustrate the influence of the substrate, we compare these results against the analytical Mie theory for an identical sphere in free-space. As seen in Fig.~\ref{fig:Ag_sph_mesh_cs}(b), the presence of the SiO\textsubscript{2} substrate breaks the symmetry, leading to a spectral red-shift and modification of the resonance compared to the free-space scenario:
\begin{lstlisting}[style=cpc_listing]
$ python3 pytools/visualization.py Ag_sphere_over_SiO2 --root sim_res --plot-csc --subjob 1 --csc scs --csc acs --csc ext --save ./sim_res/Ag_sphere_over_SiO2/out/media/cross_sections.png
$ python3 pytools/visualization.py Ag_sphere_over_SiO2 --root sim_res --plot-csc --subjob 1 --csc scs --csc acs --csc ext --mie-radius 30 --mie-material Ag --mie-background eps=1,0 --materials-root materials --mie-spline --save ./sim_res/Ag_sphere_over_SiO2/out/media/cross_sections_with_Mie.png
\end{lstlisting}

\begin{figure}[H]
\centering\includegraphics[width=0.95\columnwidth]{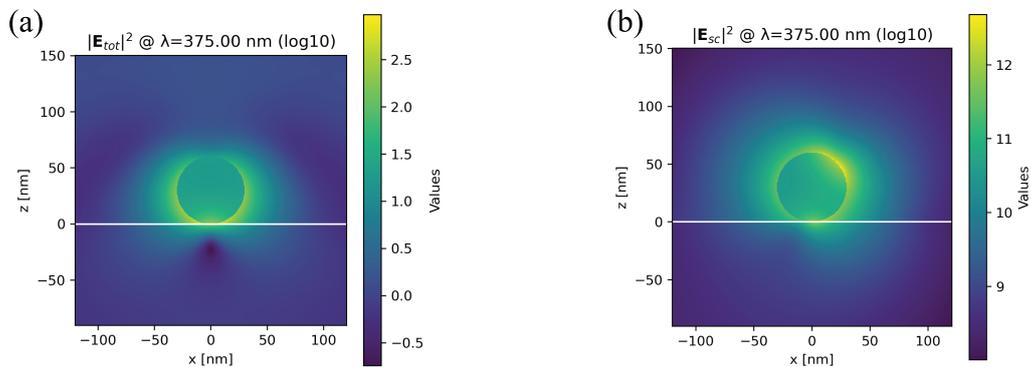}
\caption{Electric near-field intensity at $\lambda=375~\mathrm{nm}$. (a) Total field for plane wave excitation (propagation towards $-z$ and is polarization along $x$), and (b) scattered field for in-plane dipole excitation ($x = 35~\mathrm{nm},\, z = 70~\mathrm{nm},\, \theta = \pi/4,\, \varphi = \pi$).}
\label{fig:Ag_sph_near_field}
\end{figure}

Near-field intensities are computed at $\lambda=375~\mathrm{nm}$. We define a sampling grid in the $y=0$ plane and visualize, as depicted in Fig.~\ref{fig:Ag_sph_near_field}, the total electric field for a plane wave illumination that propagates towards $-z$ and is $x$-polarized, and the scattered field for a dipole illumination ($x = 35~\mathrm{nm},\, z = 70~\mathrm{nm},\, \theta = \pi/4,\, \varphi = \pi$). The post-processing and visualization commands are:
\begin{lstlisting}[style=cpc_listing]
$ python3 run_sie.py make-points Ag_sphere_over_SiO2 xz --x -120 120 --z -90 150 --step 0.6 --y0=0 -o points_xz.pos
$ python3 run_sie.py post Ag_sphere_over_SiO2 -th 0 -a -p points/points_xz.pos --lambdas 375
$ python3 pytools/visualization.py Ag_sphere_over_SiO2 --root sim_res --points points_xz.pos --field e --part tot --quantity total --function intensity --mode 2d --subjob 1 --log --hlines 0 --lambda 375 --save ./sim_res/Ag_sphere_over_SiO2/out/media/PW_375nm_Etot_intensity.png
$ python3 pytools/visualization.py Ag_sphere_over_SiO2 --root sim_res --points points_xz.pos --field e --part sc --quantity total --function intensity --mode 2d --subjob 2 --log --hlines 0 --lambda 375 --save ./sim_res/Ag_sphere_over_SiO2/out/media/dip_375nm_Esc_intensity.png
\end{lstlisting}

\subsubsection{Triangular hole in gold film}
\label{subsubsec:tri_hole}

\begin{figure}[htbp]
\centering\includegraphics[width=0.95\columnwidth]{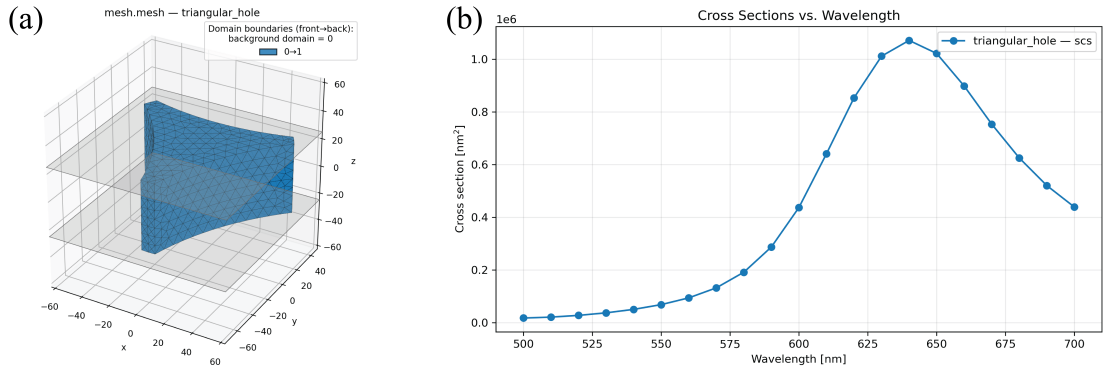}
\caption{Triangular hole in an Au film. (a) Triangular mesh (provided in \texttt{sim\_data/triangular\_hole/}) with interfaces, and (b) $C_{scs}$ for an incident plane wave that propagates towards $-z$ and is $x$-polarized.}
\label{fig:tri_hole_mesh_scs}
\end{figure}

Next, we investigate a triangular aperture in a gold film with thickness $d=50~\mathrm{nm}$. This example is particularly relevant for verifying the code implementation for bodies that are located solely within intermediate layers. The geometry is prepared and meshed, see Fig.~\ref{fig:tri_hole_mesh_scs}(a), as follows:
\begin{lstlisting}[style=cpc_listing]
$ python3 run_sie.py prepare triangular_hole --mode 2 --materials eps --spline
$ mkdir sim_res/triangular_hole/out/media/
$ python3 pytools/plot_mesh.py --sim triangular_hole --save ./sim_res/triangular_hole/out/media/mesh.png
\end{lstlisting}
After solving the problem, $C_{scs}$ is computed to identify spectral features, as in Fig.~\ref{fig:tri_hole_mesh_scs}(b):
\begin{lstlisting}[style=cpc_listing]
$ python3 run_sie.py solve triangular_hole --mode 2 -l 2 -th 0 -a
$ python3 pytools/visualization.py triangular_hole --root sim_res --plot-csc --subjob 1 --csc scs --save ./sim_res/triangular_hole/out/media/cross_sections.png
\end{lstlisting}
The calculation of all cross-sections is realized via the decomposition and integration of the Poynting vector.

Near-field maps are generated at the resonance wavelength of $\lambda = 640~\mathrm{nm}$. We sample the field in the $xy$-plane (at $z=0$) for both plane wave (propagating towards $-z$ while polarized along $x$) and dipole ($x = 0~\mathrm{nm},\, y = -30~\mathrm{nm},\, \theta = \pi/2,\, \varphi = 0$) excitations.
\begin{figure}[htbp]
\centering\includegraphics[width=0.95\columnwidth]{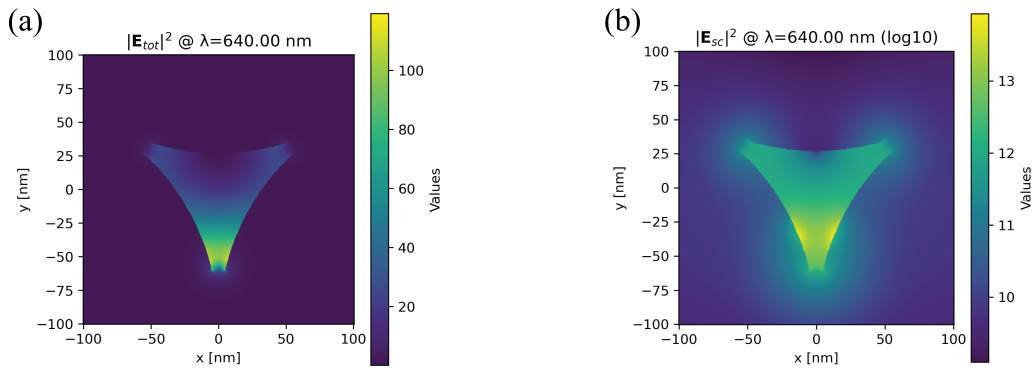}
\caption{Total and scattered electric field intensities at $\lambda=640~\mathrm{nm}$ in the $z=0$ plane for (a) a plane wave propagating towards $-z$ while polarized along $x$, and (b) a dipole illumination ($x = 0~\mathrm{nm},\, y = -30~\mathrm{nm},\, \theta = \pi/2,\, \varphi = 0$).}
\label{fig:tri_hole_near_field}
\end{figure}
The resulting total and scattered field intensity distributions, as shown in Fig.~\ref{fig:tri_hole_near_field}(a) and Fig.~\ref{fig:tri_hole_near_field}(b) respectively, demonstrate the strong field confinement at the lower sharp vertex of the aperture for plane wave illumination, and the near-field response for a dipole situated within the hole ($z=0$):
\begin{lstlisting}[style=cpc_listing]
$ python3 run_sie.py make-points triangular_hole xy --x -100 100 --y -100 100 --step 0.5 --z0=0 -o points_xy.pos
$ python3 run_sie.py post triangular_hole -th 0 -a -p points/points_xy.pos --lambdas 640
$ python3 pytools/visualization.py triangular_hole --root sim_res --points points_xy.pos --field e --part tot --quantity total --function intensity --mode 2d --subjob 1 --lambda 640 --save ./sim_res/triangular_hole/out/media/PW_640nm_Etot_intensity.png
$ python3 pytools/visualization.py triangular_hole --root sim_res --points points_xy.pos --field e --part sc --quantity total --function intensity --mode 2d --subjob 2 --log --lambda 640 --save ./sim_res/triangular_hole/out/media/dipx_640nm_Esc_intensity.png   
\end{lstlisting}

In general, when triangular elements lie on an interface of the stratified background, the tangential electromagnetic fields in the final linear system are parallel to the boundary. Due to their continuity, any side of the interface can be considered for field computation. To avoid any extra computationally expensive inter-layer calculations, the aforementioned triangles are identified as part of the layer that is closer to the scatterer's center.

\subsubsection{Cylindrical hole in multilayer stack}
\label{subsubsec:cyl_hole}

\begin{figure}[htbp]
\centering\includegraphics[width=0.95\columnwidth]{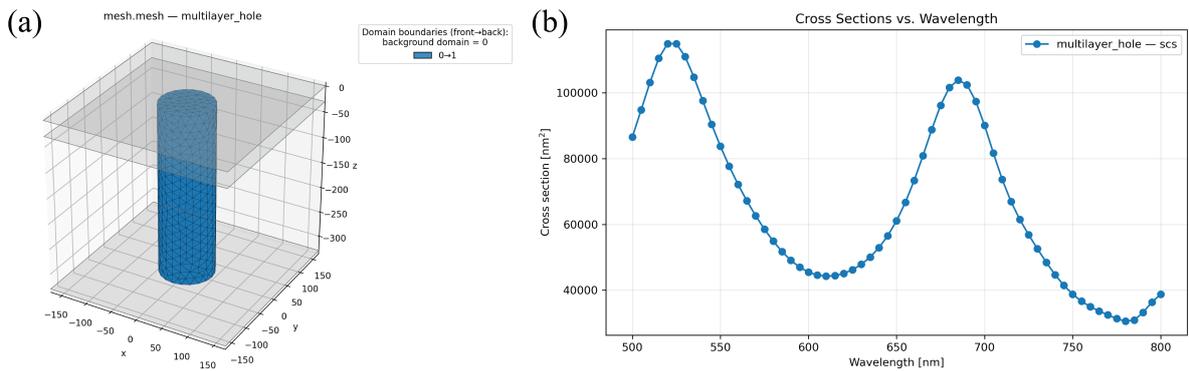}
\caption{Cylindrical hole in a multilayer stack. (a) Triangular mesh with interfaces, and (b) $C_{scs}$ for an incident plane wave that propagates towards $-z$ and is $x$-polarized. The latter shows two distinct resonances.}
\label{fig:cyl_hole_mesh_scs}
\end{figure}

For the final example, we consider a more complex stratified environment, i.e.~a cylindrical hole penetrating a multilayer stack. This scenario tests the implementation of the line integral corrections, see Eqs.~\eqref{eq:testing_line_int} and \eqref{eq:basis_line_int}, which are required when the mesh straddles through one or more interfaces between layers. Additionally, it verifies the robustness of the tabulation-interpolation scheme for Green's function, in the case of multiple layers with varying optical properties. As always, the simulation preparation and mesh inspection, see Fig.~\ref{fig:cyl_hole_mesh_scs}(a), are done first:
\begin{lstlisting}[style=cpc_listing]
$ python3 run_sie.py prepare multilayer_hole --mode 2 --materials eps --spline
$ mkdir sim_res/multilayer_hole/out/media/
$ python3 pytools/plot_mesh.py --sim multilayer_hole --save ./sim_res/multilayer_hole/out/media/mesh.png
\end{lstlisting}
The planar interfaces are located at $z=-330, -30, 0~\mathrm{nm}$. Considering the $+z$ direction, the materials of the different layers are SiO\textsubscript{2}, amorphous Si, Au, and air. Regarding the cylindrical hole, it has diameter $D=100~\mathrm{nm}$ and height $h=330~\mathrm{nm}$. Following the solution step we compute $C_{scs}$, as in Fig.~\ref{fig:cyl_hole_mesh_scs}(b), which reveals two distinct resonant wavelengths at approximately 520 nm and 685 nm:
\begin{lstlisting}[style=cpc_listing]
$ python3 run_sie.py solve multilayer_hole --mode 2 -l 2 -th 0 -a
$ python3 pytools/visualization.py multilayer_hole --root sim_res --plot-csc --subjob 1 --csc scs --save ./sim_res/multilayer_hole/out/media/cross_sections.png
\end{lstlisting}
The system is illuminated by a plane wave that propagates towards $-z$ and is $x$-polarized.

\begin{figure}[htbp]
\centering\includegraphics[width=0.95\columnwidth]{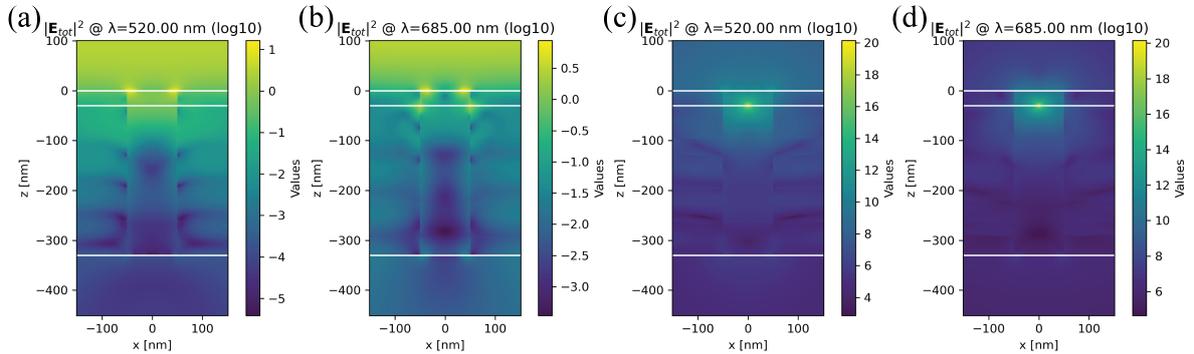}
\caption{Total electric field intensity distributions in the $xz$-plane. (a, c) Results at $\lambda=520~\mathrm{nm}$, and (b, d) at $\lambda=685~\mathrm{nm}$ for plane wave and dipole illuminations. The incident plane wave propagates towards $-z$ and is $x$-polarized, while the dipole is located at ($x=y=0~\mathrm{nm},\, z=-30~\mathrm{nm}$) with direction ($\theta=\pi/2,\, \varphi=0$). The horizontal lines indicate the interfaces of the multilayer stack.}
\label{fig:cyl_hole_near_field}
\end{figure}

Near-field distributions are analyzed in the $xz$-plane at the two resonant wavelengths:
\begin{lstlisting}[style=cpc_listing]
$ python3 run_sie.py make-points multilayer_hole xz --x -150 150 --z -450 100 --step 1.5 --y0=0 -o points_xz.pos
$ python3 run_sie.py post multilayer_hole -th 0 -a -p points/points_xz.pos --lambdas 520
$ python3 run_sie.py post multilayer_hole -th 0 -a -p points/points_xz.pos --lambdas 685
\end{lstlisting}
To clearly visualize the intensity profiles relative to the layer boundaries, the \texttt{----hlines} argument is used to overlay the interface positions in Fig.~\ref{fig:cyl_hole_near_field}. For plane wave and dipole illuminations at $\lambda = 520~\mathrm{nm}$ the commands are:
\begin{lstlisting}[style=cpc_listing]
$ python3 pytools/visualization.py multilayer_hole --root sim_res --points points_xz.pos --field e --part tot --quantity total --function intensity --mode 2d --subjob 1 --log --hlines=-330,-30,0 --lambda 520 --save ./sim_res/multilayer_hole/out/media/PW_520nm_Etot_intensity.png
$ python3 pytools/visualization.py multilayer_hole --root sim_res --points points_xz.pos --field e --part tot --quantity total --function intensity --mode 2d --subjob 2 --log --hlines=-330,-30,0 --lambda 520 --save ./sim_res/multilayer_hole/out/media/dipx_520nm_Etot_intensity.png
\end{lstlisting}
The same distributions are extracted also for $\lambda = 685~\mathrm{nm}$:
\begin{lstlisting}[style=cpc_listing]
$ python3 pytools/visualization.py multilayer_hole --root sim_res --points points_xz.pos --field e --part tot --quantity total --function intensity --mode 2d --subjob 1 --log --hlines=-330,-30,0 --lambda 685 --save ./sim_res/multilayer_hole/out/media/PW_685nm_Etot_intensity.png
$ python3 pytools/visualization.py multilayer_hole --root sim_res --points points_xz.pos --field e --part tot --quantity total --function intensity --mode 2d --subjob 2 --log --hlines=-330,-30,0 --lambda 685 --save ./sim_res/multilayer_hole/out/media/dipx_685nm_Etot_intensity.png    
\end{lstlisting}

\begin{figure}[htbp]
\centering
  \includegraphics[width=0.95\columnwidth]{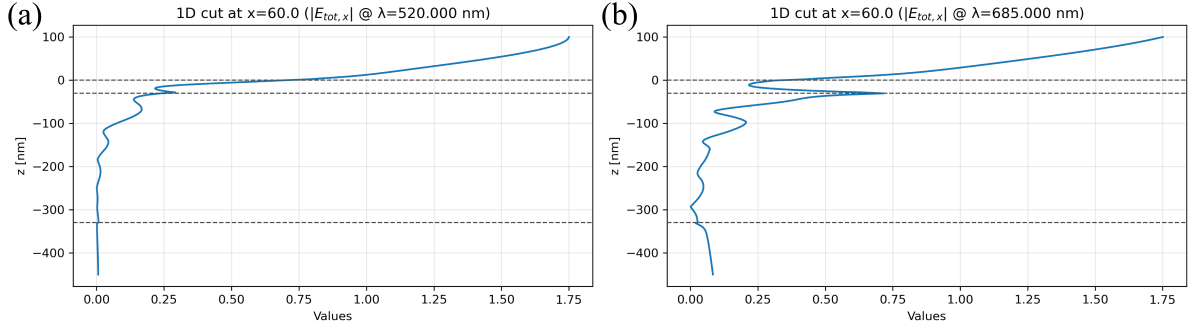}
  \caption{Vertical line profiles of the tangential electric field magnitude $|E_{tot,x}|$ at $x=60~\mathrm{nm}$. The profiles for (a) $\lambda = 520~\mathrm{nm}$ and (b) $\lambda = 685~\mathrm{nm}$ demonstrate field continuity across all interfaces (dashed lines).}
  \label{fig:line_near-field_pw}
\end{figure}

The continuity of tangential electromagnetic fields is verified in Fig.~\ref{fig:line_near-field_pw}, where the magnitude of the $x$-component of the total electric field is plotted along a vertical line cut of the $y=0$ plane, at $x = 60~\mathrm{nm}$. Since the incident field is $x$-polarized and all interfaces are perpendicular to the $z$-axis, $E_{tot,x}$ represents the tangential component of the total electric field, which must remain continuous across all boundaries. As shown in both panels, the field profiles exhibit perfect continuity at the interfaces, confirming the validity of the line integral implementation in our model:
\begin{lstlisting}[style=cpc_listing]
$ python3 pytools/visualization.py multilayer_hole --root sim_res --points points_xz.pos --field e --part tot --quantity x --function magnitude --mode 1d --subjob 1 --line x=60 --hlines=-330,-30,0 --lambda 520 --save ./sim_res/multilayer_hole/out/media/abs_Etot_x_520nm_1d.png
$ python3 pytools/visualization.py multilayer_hole --root sim_res --points points_xz.pos --field e --part tot --quantity x --function magnitude --mode 1d --subjob 1 --line x=60 --hlines=-330,-30,0 --lambda 685 --save ./sim_res/multilayer_hole/out/media/abs_Etot_x_685nm_1d.png
\end{lstlisting}
Finally, the calculations in this scattering problem are computationally more expensive compared to previous stratified media examples, due to inter-layer interactions and 3D tabulation-interpolation.

\section{Conclusion}
\label{sec:conclusion}

In this paper, we have presented \textls[120]{HELIOS}, a versatile open-source software package based on the SIE method for simulating electromagnetic scattering by nanostructures in homogeneous, periodic and stratified environments. By implementing the PMCHWT formulation, the solver provides stable and accurate solutions for arbitrary 3D geometries.

A key feature of \textls[120]{HELIOS} is its comprehensive handling of diverse scenarios. For periodic systems, the integration of Ewald's transformation enables the efficient evaluation of periodic Green's tensors, facilitating the rigorous analysis of photonic crystals and metasurfaces. Furthermore, the code addresses the complexities of stratified media through the matrix-friendly formulation. The combination of advanced singularity cancellation techniques, optimized Sommerfeld integration paths, and a tabulation-interpolation scheme ensures both numerical precision and computational efficiency, even for structures penetrating multiple interfaces.

The software architecture leverages a C\texttt{++} core for intensive matrix operations, coupled with a user-friendly Python interface that streamlines the simulation workflow, from pre-processing to visualization. The accuracy and flexibility of the code have been demonstrated through a series of examples, including isolated nanoparticles, periodic arrays, and apertures in multilayer stacks. These tests not only validate the numerical implementation but also highlight the code's capability to extract key optical quantities, such as cross-sections, near-field distributions, and scattering parameters.

We believe that \textls[120]{HELIOS} constitutes a valuable resource for the research community, offering a reliable tool for the design and analysis of complex nanophotonic devices. Its open-source nature encourages collaborative developments, paving the way for future extensions.

\section*{CRediT authorship contribution statement}

\textbf{Parmenion S.~Mavrikakis:} Conceptualization, Formal analysis, Investigation, Methodology, Software, Validation, Visualization, Writing - original draft, Writing – review and editing. \textbf{Olivier J.~F.~Martin:} Conceptualization, Funding acquisition, Project administration, Resources, Supervision, Writing – review and editing.


\section*{Declaration of competing interest}

The authors declare that they have no known competing financial interests or personal relationships that could have appeared to influence the work reported in this paper.

\section*{Data availability}

Data will be made available on request.

\section*{Acknowledgments}

It is a pleasure to acknowledge contributions from A.M.~Kern (free-space scattering) and B.~Gallinet (periodic systems) in the initial development of \textls[120]{HELIOS}. This work was supported by the Swiss National Science Foundation (SNSF) under project 200021 212758.







\end{document}